\documentclass[smallextended]{svjour3}

\usepackage{latexsym}
\usepackage{amssymb}
\usepackage{amsmath}
\usepackage{graphicx}
\usepackage{multirow}
\newcommand{\N}{{\rm I}\!{\rm N}}
\newcommand{\R}{{\rm I}\!{\rm R}}
\DeclareMathOperator*{\argmin}{argmin}
\usepackage{color}

\usepackage{lipsum}
\usepackage{amsfonts}
\usepackage{epstopdf}
\usepackage{algorithmic}

\ifpdf
\DeclareGraphicsExtensions{.eps,.pdf,.png,.jpg}
\else
\DeclareGraphicsExtensions{.eps}
\fi

\begin{document}
	
\title{A micro-to-macro approach to returns, volumes and waiting times}

\author{Guglielmo D'Amico
	\and Filippo Petroni}

\institute{G. D'Amico \at
	Department of Pharmacy, Universit\`a `G. d'Annunzio' di Chieti-Pescara, Italy\\
	\email{g.damico@unich.it}           
	\and
	 F. Petroni \at
	Department of Management, Universit\`a Politecnica delle Marche, Ancona, Italy
	\email{f.petroni@univpm.it}   
}

\maketitle


\begin{abstract}
Fundamental variables in financial market are not only price and return but a very important role is also played by trading volumes.
Here we propose a new multivariate model that takes into account price returns, logarithmic variation of trading volumes and also waiting times, the latter to be intended as the time interval between changes in trades, price, and volume of stocks.
Our approach is based on a generalization of semi-Markov chains where an endogenous index process is introduced. We also take into account the dependence structure between the above mentioned variables by means of copulae. The proposed model is motivated by empirical evidences which are known in financial literature and that are also confirmed in this work by analysing real data from Italian stock market in the period August 2015 - August 2017. By using Monte Carlo simulations, we show that the model reproduces all these empirical evidences. 
\end{abstract}




\section{Introduction}
In financial markets, high frequency data and modelling have acquired a dominant role due to relevant information brought by intra-day observations. Nowadays, even more sophisticated models and ideas can be advanced and tested on real market data based on the huge amount of information that can be stored and processed on modern computers.\\
\indent A large part of effort in market microstructure studies has been produced in order to understand, mimic and predict basic empirical regularities observed in the most important financial variables. A special attention has been dedicated to the relation between financial volumes and returns. The majority of the works in this area can be classified within the so-called econometric framework, sometimes also referred as macro-to-micro approach. The cornerstone of this approach is to consider the observed price to be a collateral effect of an unobservable volatility process to which a noise process transformation is applied, see e.g. \cite{boll86}. This line of research has flourished during the last two decades and considerable attention has been dedicated to the problem of irregular spacing in time of observations when dealing with high frequency financial data; the seminal work by \cite{engl00} and the recent review by \cite{bhog19} can provide a wide overview on the subject. Rapidly, econometricians turned the attention on multivariate models of logarithmic price returns, volumes and duration (waiting times), the latter to be intended as the time interval between changes in trades, price, and volume of stocks, see e.g. \cite{jain88,mang05,dejo09,podo09}\\
\indent Another strand of literature relies on the modelling of directly observable quantities, the so-called micro-to-macro approach that is philosophically in contrast with the econometric approach. This framework has a long tradition that has its roots in the paper by \cite{bach00} and also embraces lattice based models including the popular binomial and trinomial models (see e.g. \cite{cox79} and \cite{boyl86}). The micro-to-macro approach has undergone a revival in recent years mainly due to the work of econophysicists that introduced the Continuous Time Random Walks (CTRW) apparatus in the modelling of financial returns, see \cite{mai00,rab02,repe04}. The evolution equation of CTRW was formulated and it was shown that it can catch non-Markovian effects. Sometimes the non-Markovian behaviour of stocks has been accommodated considering a latent Markov process acting as a switching process as done in \cite{cura15}. In any case, a viable solution to non-Markovian problem is given by semi-Markov based models. Semi-Markov processes are the equivalent of CTRW having a non-independent space-time dynamic. They appeared in the fifties in the probability field due to the independent contributions by \cite{levy56} and \cite{smit55}. They have been successfully investigated and applied in connection with a very wide range of problems including reliability theory, queuing, stochastic systems and DNA analysis, see e.g. \cite{jans06}, \cite{barb08}, \cite{silv17,silv18} and \cite{das10,das16}.\\
\indent Financial modeling is not an exception and has assisted to the progressive abandonment of the Markovian property in favour of the semi-Markov one. Examples comes from credit rating modelling (e.g. \cite{dami05,vasi06,vasi13}), high-frequency financial data (\cite{swis11,fodr15,swis17}), financial time series (\cite{bull06,nyst15}) and pricing problems (\cite{dami09,silv04,swis13}).\\
\indent However, only recently it has been recognized that also semi-Markov processes (CTRW included) are not able to reproduce accurately the statistical properties of high-frequency financial data and a more general solution has been advanced in a series of papers where the concept of weighted-indexed semi-Markov chains (WISMC) has appeared, see \cite{dape11} and \cite{dape12}. The WISMC model represents a generalization of ordinary semi-Markov processes and revealed to be particular useful to reproduce long-term dependence in the stock returns among other stylized facts of statistical finance. WISMC models were extended using different strategies to multivariate settings and applied to measure risk of financial portfolios, see \cite{dape14} and \cite{dape_18}. In the meantime they were also successfully applied to the modelling of financial volumes by \cite{dape18b}.\\
\indent So far, models proposed in the literature of the micro-to-macro approach have not yet been able to advance a unifying approach where returns, volumes and waiting times are jointly modelled in such a way to reproduce known empirical regularities they possess. The contribution of this paper is to present a modelling framework where these three variables are managed contemporaneously in a satisfactory and flexible way. In particular, we first conduct a detailed explorative data analysis with the aim of a better understanding of the empirical relationships among the considered financial variables. The data analysed are from four Italian stocks for the period August 2015 - August 2017 observed at 1 minute frequency. The WISMC model is presented for the first time ever in discrete time with a general state space and is considered as a model for the log-returns and also for the log-volume returns with different kernels. To achieve the objective of a multivariate model of price-volumes-waiting times (triplet process), specific data-driven assumptions are advanced and the dependence structure between the price and volume return processes conditional on the waiting time process is embodied by using a copula function on the joint distribution of modulus of returns and modulus of volume returns that exhibit a general dependence structure. The dynamic of the multivariate model is completely characterized by the determination of the kernel of the triplet process. In general, this allows the computation of any financial statistic that can be written as a functional of the kernel of the triplet process. The model is used to compute linear and nonlinear measures of dependence, joint first passage time distribution function of price and volumes and shows ability to reproduce probability density functions of both variables as well as cross and auto-correlation functions.\\
\indent The paper is organized as follows. Section 2 provides a statistical analysis of financial data with a particular focus on the relationships among price returns, volume returns and waiting times. Section 3 sets out the marginal models of price and volumes and the multivariate extension by means of copula functions. In this section the kernel of the triplet process is studied under some assumptions that are justified by the data. The section presents also the computation of some financial functions of broad interest. Section 4 illustrates the result of the application to real data and demonstrates the accuracy of the model in reproducing the main empirical regularities observed in financial markets. Section 5 summarizes our contribution and results. All proofs are deferred to the “Appendix”.

\section{Data analysis}\label{database}
\begin{figure}
	\centering
	\includegraphics[width=8cm]{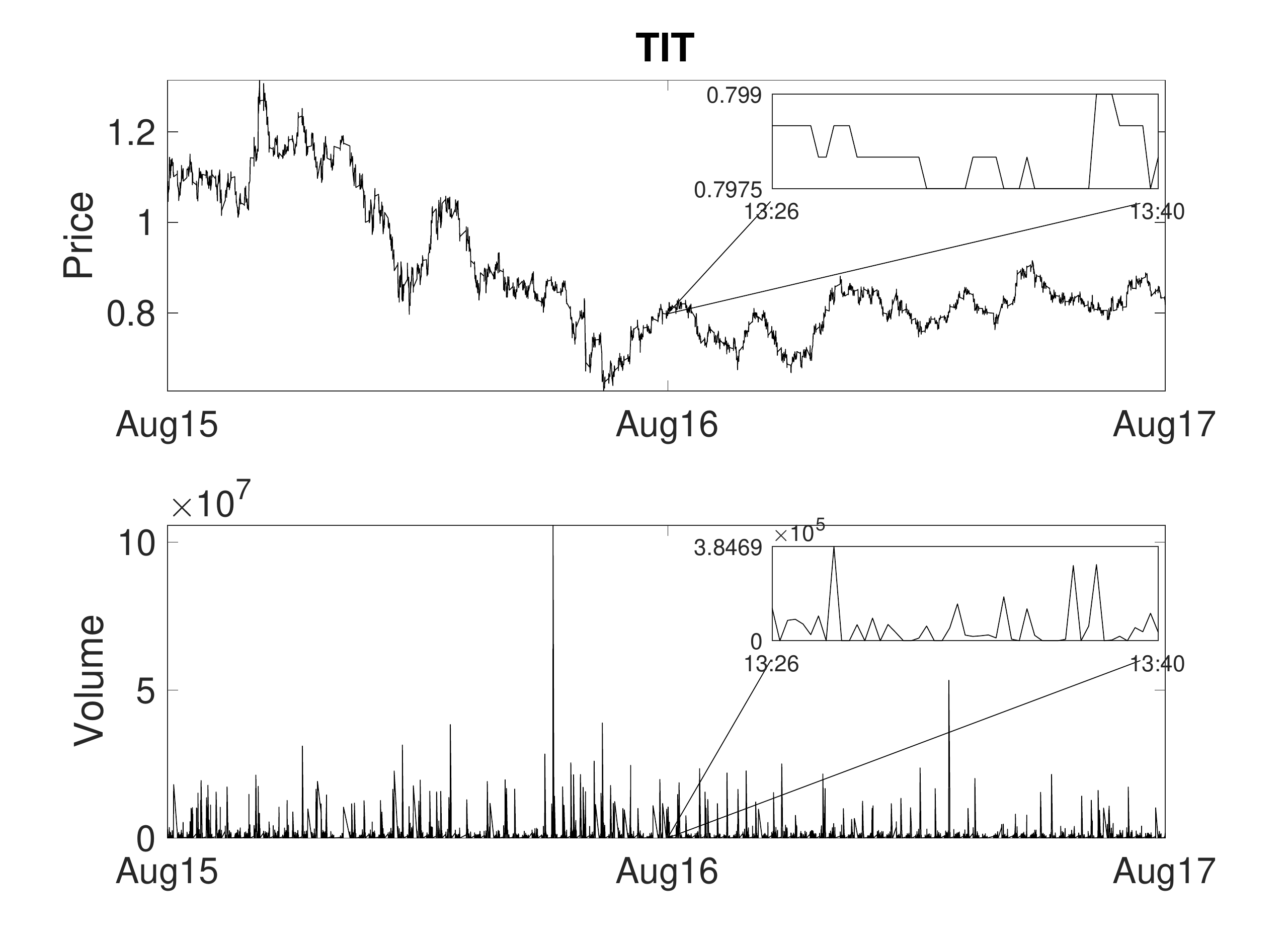}
	\caption{Time series of prices and volumes for stock TIT.} \label{price_vol}
\end{figure}

Empirical research on price changes, financial volumes and waiting times has identified some characteristics often called the stylized facts \cite{guil97,pas00,pas99,pas99b,bav01,pet16,pet03}. In this section we conduct an explorative data analysis with the aim of a better understanding of empirical relations among the considered financial variables. This analysis will inspire the main assumptions under which our model is going to be built on in the next section.\\
\indent The data used are quotes of Italian stocks  for the period August 2015 - August 2017 (2 full years) with 1 minute frequency. Every minute, the last price and the cumulated volume (number of transactions) is recorded. For each stock the database
is composed of about $2.6*10^5$ volumes and prices. The list of stocks analysed and their symbols are reported in Table \ref{tab}. 
\begin{table}
	\begin{center}
		\begin{tabular}{|l|l|}
			\hline			
			\textbf{Code} & \textbf{Name}\\\hline
			\textbf{TIT} & Telecom\\\hline
			\textbf{ISP} & Intesa San Paolo\\\hline
			\textbf{TEN} & Tenaris \\\hline	
			\textbf{F} & FCA Group\\\hline	
		\end{tabular}
	\end{center}
	\caption{Stocks used in the application and their symbols}\label{tab}
\end{table}
From now onward we will use only the codes in the table to identify each stock.\\
\indent The analysed stocks are chosen to represent different market sectors. According to the Global Industry Classification Standard \textbf{F} is in the industrial sector, \textbf{ISP} is one of the largest banks in Italy (financial sector), \textbf{TIT} is in the telecommunication and \textbf{TEN} in the energy sector.
In Figure \ref{price_vol} we show an example, for the stock \textbf{TIT}, of the time series of price {$S(t)$} and trading volumes {$V(t)$} in the analysed period.  


As a first step we analyse the time series and look for the most important statistical features. From prices we build a time series of the price returns defined as $r(t) = \log(S(t)/S(t-1))$ and from trading volumes we define the log variation (from now onward volume returns) as $v(t) = \log(V(t)/V(t-1))$ where $t$ is the time variable in one minute frequency. To be sure to use only variation in one minute period we exclude from the analyses the variation of both variables from the closing of the stock market at day $d$ to the re-opening in the next trading day $d+1$ (we remind that the stock market is open from 9 am to 17:30 pm in weeks day).

\begin{figure}
	\centering
	\includegraphics[width=8cm]{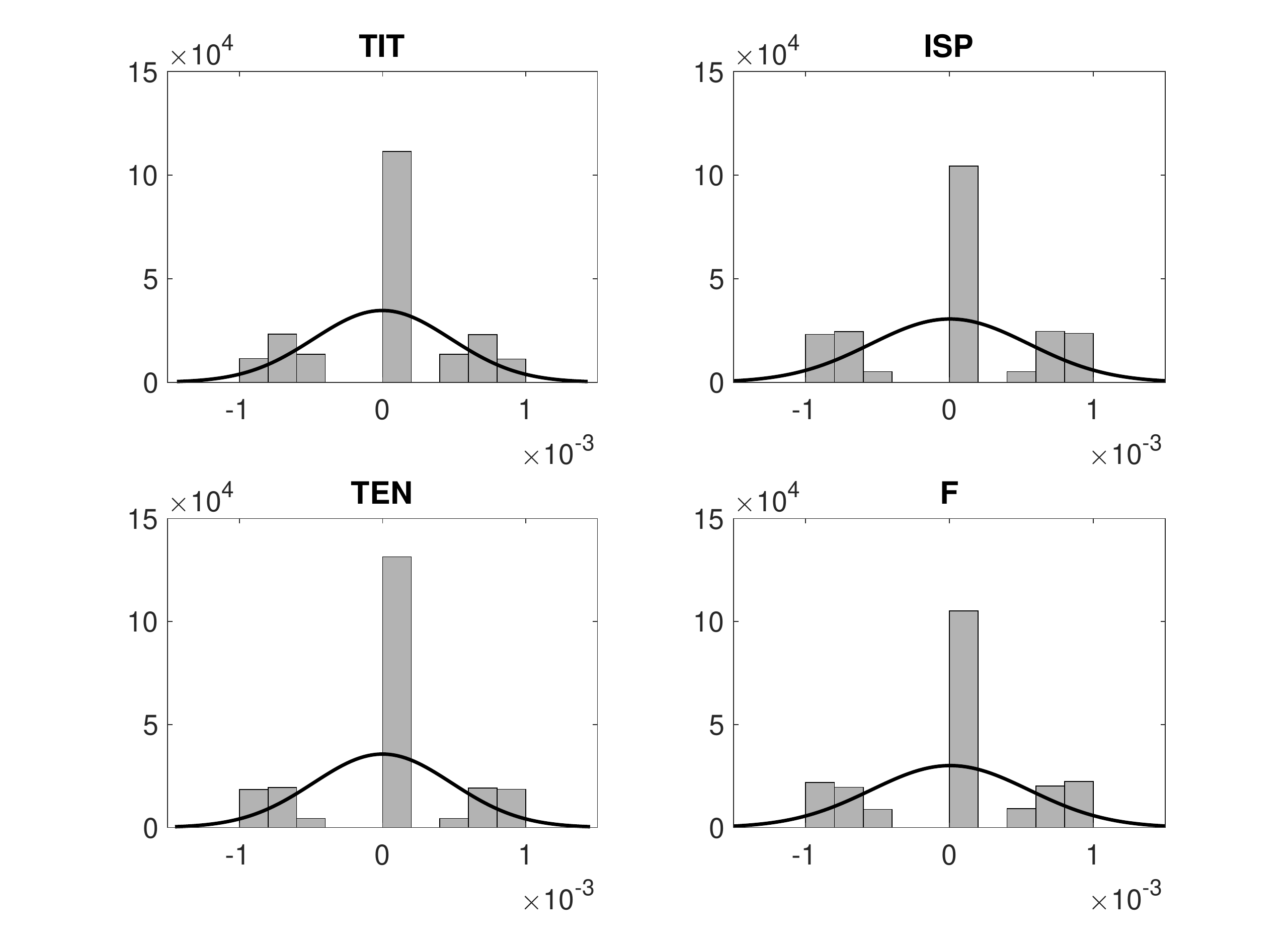}
	\caption{Histogram of $r(t)$ compared with a Gaussian fit.} \label{prob_rit}
\end{figure}

\begin{figure}
	\centering
	\includegraphics[width=8cm]{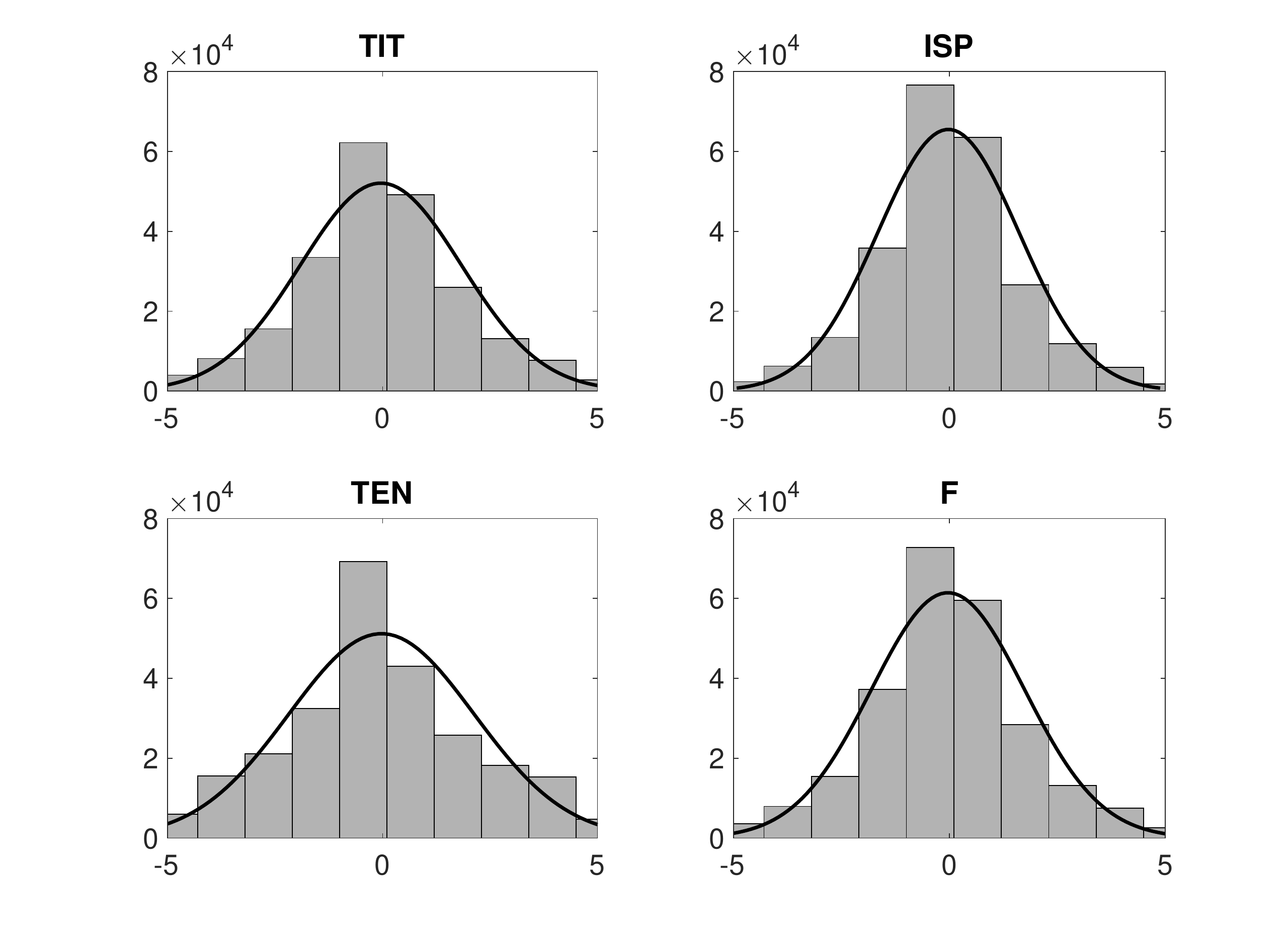}
	\caption{Histogram of $v(t)$ compared with a Gaussian fit.} \label{prob_vol}
\end{figure}

In Table \ref{table_descr_rit} we summarize the descriptive statistics of price returns $r(t)$,
while in Table \ref{table_descr_vol} we summarize the descriptive statistics of volume returns $v(t)$.
\begin{table}
	\begin{center}
		\scalebox{0.8}{
		\begin{tabular}{|l|c|c|c|c|c|}
			\hline\noalign{\smallskip}
			\textbf{Stock} & \textbf{Mean}& \textbf{Median}& \textbf{Standard deviation}& \textbf{Skewness}& \textbf{Kurtosis}\\
			\noalign{\smallskip}\hline
				\textbf{TIT}&$-2*10^{-6}$\ &$0$\ &$4.7*10^{-4}$\ &$0.18*10^{-2}$&$2.5$\\\hline	
			\textbf{ISP}&$2.5*10^{-6}$\ &$0$\ &$5.4*10^{-4}$\ &$-0.25*10^{-2}$&$2.2$\\\hline
			\textbf{TEN}&$-4.2*10^{-7}$\ &$0$\ &$4.8*10^{-4}$\ &$0.24*10^{-3}$&$2.8$\\\hline
				\textbf{F}&$4.7*10^{-6}$\ &$0$\ &$5.5*10^{-4}$\ &$-0.88*10^{-2}$&$2.3$\\\hline
		\end{tabular}}
	\end{center}
	\caption{Descriptive statistics of the price returns $r(t)$.}\label{table_descr_rit}
\end{table}
\begin{table}
	\begin{center}
		\scalebox{0.8}{
		\begin{tabular}{|l|c|c|c|c|c|}
			\hline\noalign{\smallskip}
			\textbf{Stock} & \textbf{Mean}& \textbf{Median}& \textbf{Standard deviation}& \textbf{Skewness}& \textbf{Kurtosis}\\
			\noalign{\smallskip}\hline
			\textbf{TIT}&$-2.9*10^{-2}$\ &$-0.06$\ &$1.58$\ &$6.6*10^{-2}$&$3.94$\\\hline	
		    \textbf{ISP}&$-2.2*10^{-2}$\ &$-0.05$\ &$1.53$\ &$9.2*10^{-2}$&$3.96$\\\hline
			\textbf{TEN}&$-2.9*10^{-2}$\ &$-0.05$\ &$1.31$\ &$5.9*10^{-3}$&$4.08$\\\hline		
			\textbf{F}&$-2.3*10^{-2}$\ &$-0.06$\ &$1.51$\ &$8.7*10^{-2}$&$3.88$\\\hline
		\end{tabular}}
	\end{center}
	\caption{Descriptive statistics of volume returns $v(t)$.}\label{table_descr_vol}
\end{table}
To better visualize the distributions of both time series we show in Figure \ref{prob_rit} and \ref{prob_vol} the histogram of $r(t)$ and $v(t)$, respectively, and we also compare them with the best Gaussian fit.

\begin{figure}
	\centering
	\includegraphics[width=8cm]{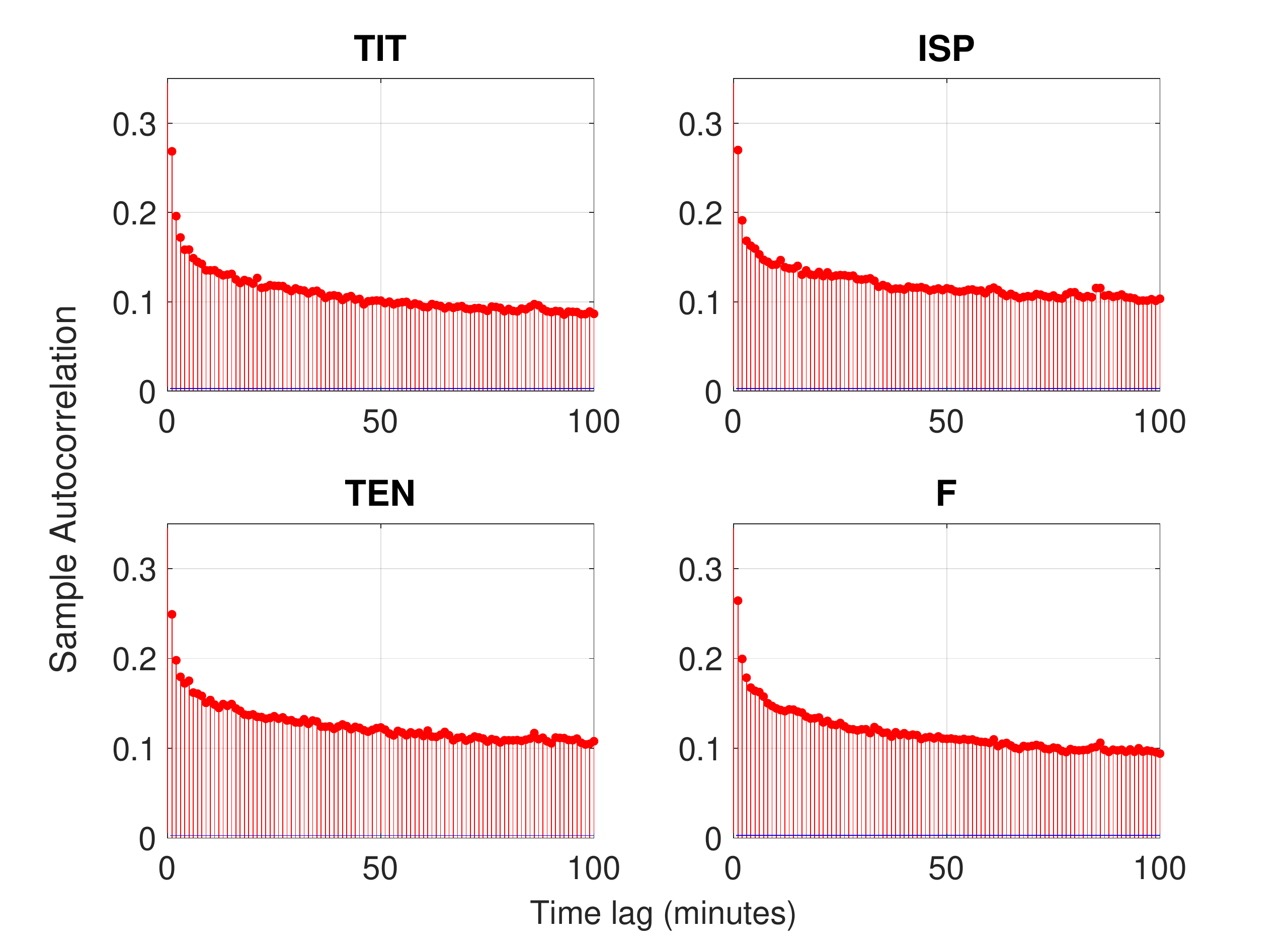}
	\caption{Sample autocorrelation of $|r(t)|$.} \label{cor_rit}
\end{figure}

\begin{figure}
	\centering
	\includegraphics[width=8cm]{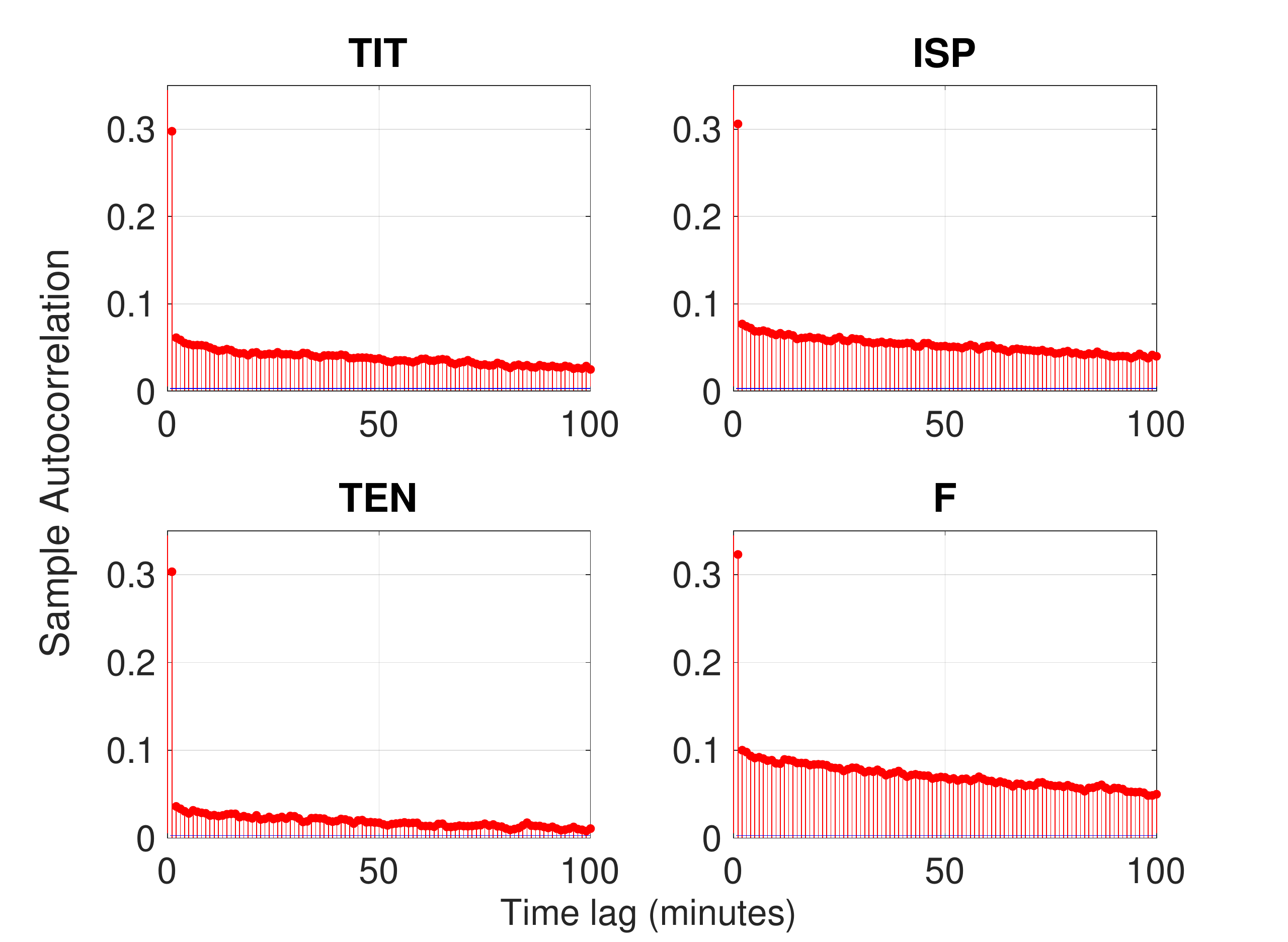}
	\caption{Sample autocorrelation of $|v(t)|$.} \label{cor_vol}
\end{figure}

We performed a Jarque-Bera test that rejected the Gaussian distribution for both $r(t)$ and $v(t)$ at $1\%$ significance level. \\
\indent One of the most important statistical feature of both time series is that their absolute values are long range correlated. We show this in Figures \ref{cor_rit} and  \ref{cor_vol}. We also found that there is zero correlation between $r(t)$ and $v(t)$ while a non-zero correlation between  $|r(t)|$ and $v(t)$ is present. This result is shown in Table \ref{table_cross_corr}.  In this Table we show all possible combination of correlation between $r(t)$ and $v(t)$ and their absolute values, we also show the p-values which gives statistical significance of non-zero correlation. 

\begin{table}
	\begin{center}
		\scalebox{0.8}{
\begin{tabular}{|l|l|l|l|l|l|l|l|l|}
	\hline
		&$\rho(r(t),v(t))$&p-value&$\rho(|r(t)|,v(t))$&p-value& $\rho(r(t),|v(t))|$&p-value&$\rho(|r(t)|,|v(t))|$&p-value\\\hline
	TIT&0.0089&0&0.086&0&-0.0032&0.023&0.020&0\\\hline
	ISP&-0.010&0&0.086&0&-0.0036&0.0095&-0.019&0\\\hline
	TEN&-0.0091&0&0.086&0&0.00055&0.69&0.040&0\\\hline
	F&-0.012&0&0.11&0&-0.0041&0.0037&-0.027&0\\\hline
	\end{tabular}}
	\end{center}
\caption{Cross correlation between price and volume returns.}\label{table_cross_corr}
\end{table}

\begin{figure}
	\centering
	\includegraphics[width=8cm]{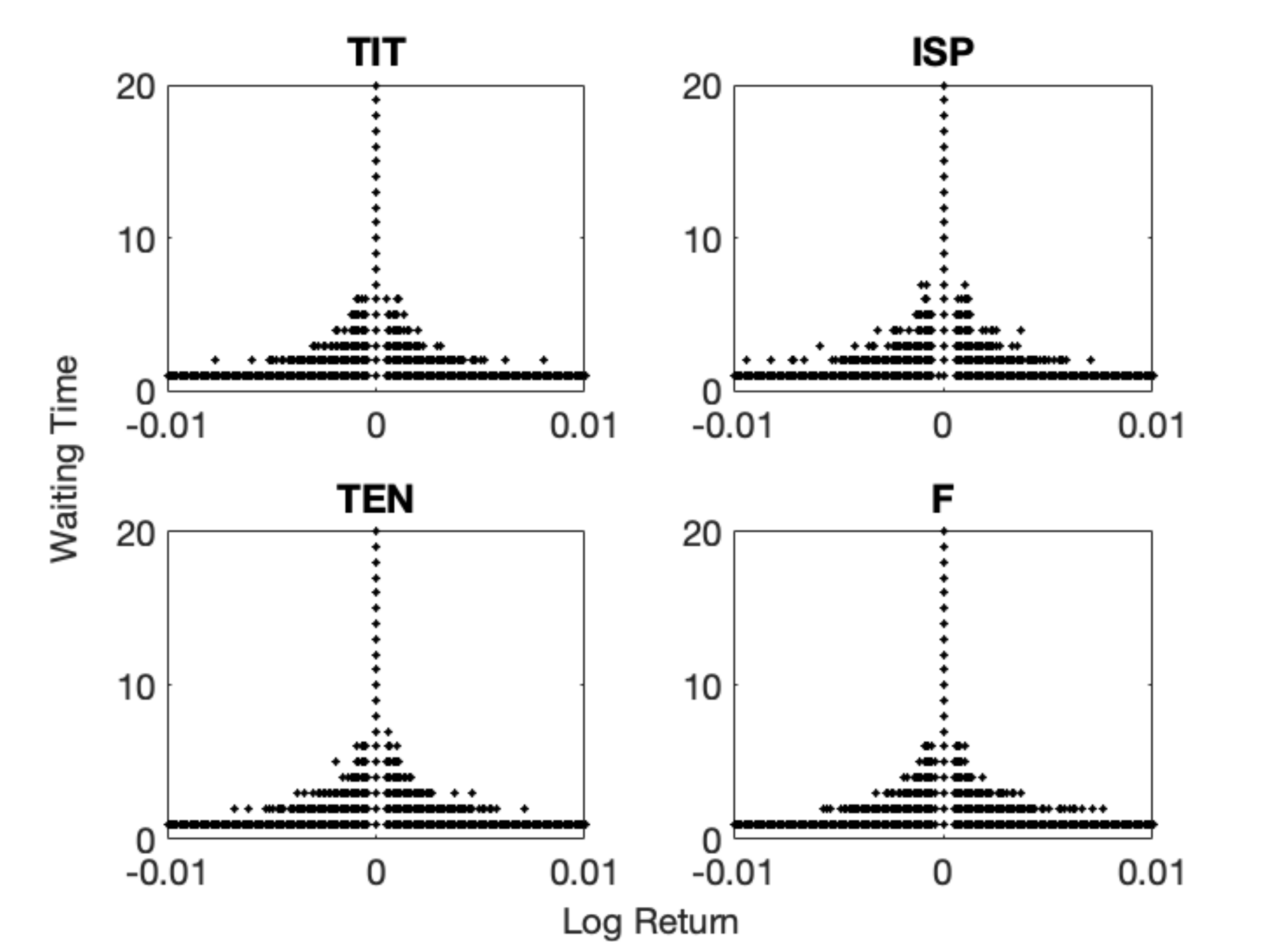}
	\caption{Waiting time as function of $r(t)$.} \label{Waitingtime_rit}
\end{figure}

\begin{figure}
	\centering
	\includegraphics[width=8cm]{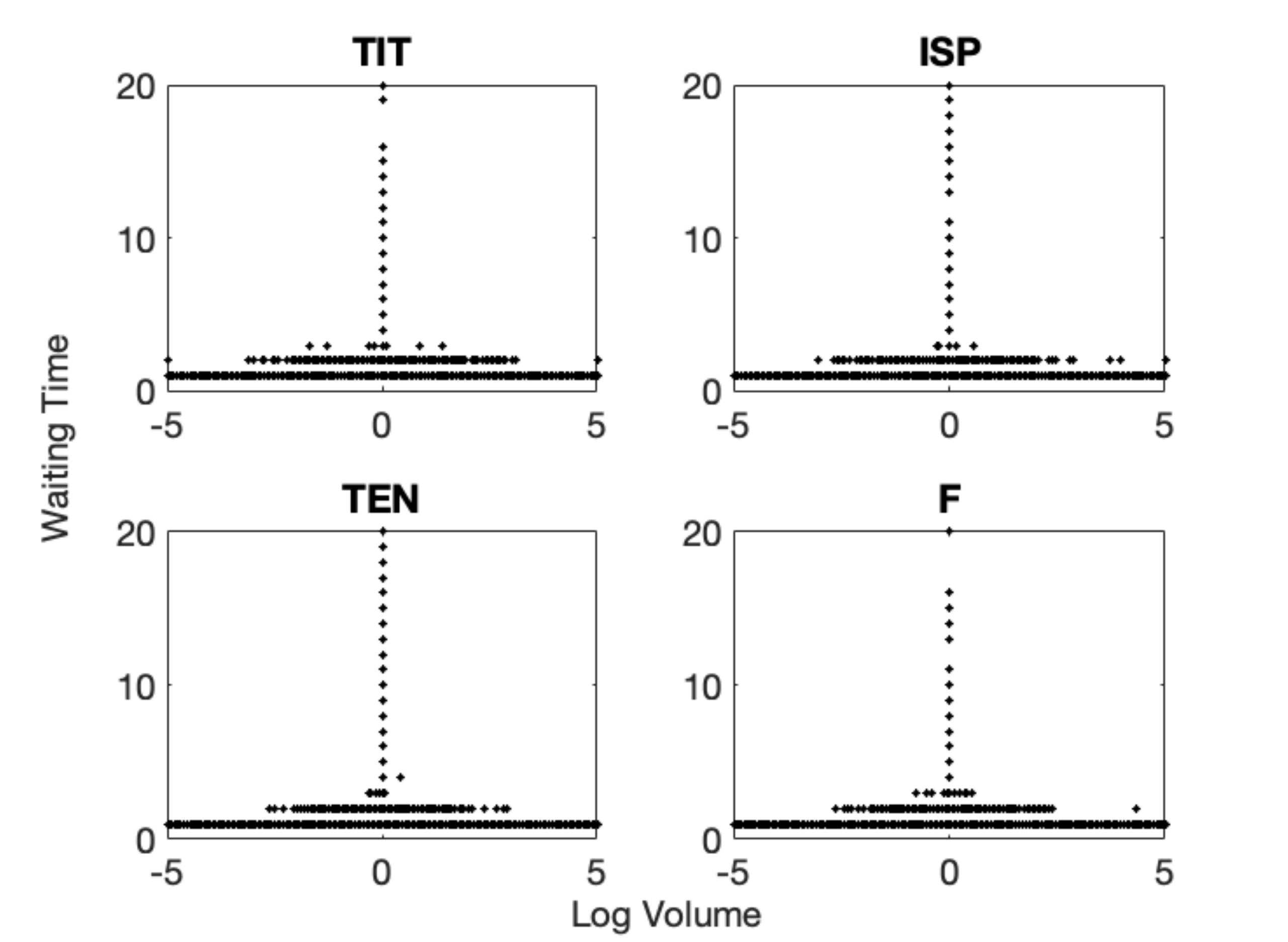}
	\caption{Waiting as function of $v(t)$.} \label{Waitingtime_vol}
\end{figure}

Given these properties a good model should be able to take all of them into account.
Another property that we found quite interesting and that should be included into a model is the following: for both time series ($r(t)$ and $v(t)$) we found that there is a dependence with the waiting time which is defined as the time it takes for returns to change their values. This can be seen in Figures \ref{Waitingtime_rit} and  \ref{Waitingtime_vol} where we have plotted the times it takes from a specific values of $r(t)$ ($v(t)$) to jump into all other values. It can be noticed that waiting times has some dependence from $r(t)$ ($v(t)$) values. This empirical evidence was already observed for price returns in \cite{mai00} and now we highlight it also for volume returns.

This result is confirmed from contingency tables where we can see the dependence between $r(t)$, $v(t)$ and the waiting time $T$.  The contingency tables, shown in Table \ref{cont_table}, have been obtained by a discretization of both $r(t)$ and $v(t)$ into 5 states.

\begin{table}
	\begin{center}
		\scalebox{0.75}{
			\begin{tabular}{|l|l|l|l|l|l|l|}
				\hline
				&&\multicolumn{5}{|c|}{$r(t)$} \\
				\cline{3-7}
				&&$-\infty : -0.13\%$&$-0.13\% : -0.05\%$&$-0.05\% : 0.05\%$&$0.05\% : 0.13\%$&$0.13\%:+\infty$\\
				\hline
				\multirow{3}{*}{$T$}&$0:2$&24647 (23386,7)&45762 (43588,8)&43459 (50186,2) &46381 (44216,1)&23984 (22855,1)\\\cline{2-7}
				&$2:4$&696 (1436,3)& 1487 (2677,1)& 6805 (3082,3)& 1553 (2715,6)&774 (1403,7)\\\cline{2-7}
				&$4:+\infty$&45 (565,0)&	70 (1053,1)	&4217 (1212,5)	&66 (1068,2)	&53 (552,2)\\\hline
		\end{tabular}}

\vspace{1cm}
\scalebox{0.8}{
			\begin{tabular}{|l|l|l|l|l|l|l|}
			\hline
			&&\multicolumn{5}{|c|}{$v(t)$} \\\cline{3-7}
			&&$-\infty:-4$&$-4:-1.3$&$-1.3:1.3$&$1.3:4$&$4:+\infty$\\\hline
			\multirow{3}{*}{$T$}&$0:2$&3784 (3071,1)&58063 (47168,3)&36466 (60051,0)&59221 (48145,5)&4787 (3885,2)\\\cline{2-7}
			&$2:4$&0 (289,7)&54 (4449,2)&15157 (5664,3)&100	(4541,3)&0 (366,5)\\\cline{2-7}
			&$4:+\infty$&0 (423,2)&0 (6499,5)&22367 (8274,7)&0 (6634,2)&0 (535,4)\\\hline
		\end{tabular}}
	\end{center}
	\caption{Contingency table for $F$. In the first table we tested the dependence between $r(t)$ and $T$ while in the second table we tested the dependence between $v(t)$ and $T$. The numbers in brackets are obtained under the independence hypotheses. The ${\chi}^{2}$ test rejects the null hypothesis of independence.}\label{cont_table}
\end{table}

It can be easily noticed, from Tables \ref{cont_table}, that there is a dependence of the number of transition from $T$ (for $T=1$ there are much more transitions than for $T=2$ or $T=3$). More specifically, in brackets we give the number of transitions that we would expect for independent processes where the probability of finding a given number of transition is simply given by the product of the frequencies of having each variable at that given state. From Tables \ref{cont_table} it is obvious that the independent hypothesis does not hold for both processes.
We obtained similar results for all other stocks which, for reasons of space, are not shown here. From the above tables and from all results obtained in this section, we can say that a good real world model of price and volumes should take into account all the afore detected stylized facts that we can summarize in a list:\\ 
- distributions of price returns and volume returns are not Gaussian;\\
- the absolute values of price returns are long range correlated;\\
- the absolute values of volume returns are long range correlated;\\
- price returns and volume returns are uncorrelated while a non-zero correlation between $|r(t)|$ and $v(t)$ is present;\\
- $r(t)$ and the waiting times influence each other;\\
- $v(t)$ and the waiting times influence each other.\\
The majority of them was already known and extensively documented in the financial literature, here they have been confirmed in our dataset. A very interesting summary and financial implications of those empirical regularities are discussed in \cite{bouc02} and in the references therein. The empirical evidences in the list are the cornerstones on which is built the model we are going to present in next section.

\section{Mathematical Model} \label{model}

In this section we first present the WISMC model that is used as a marginal model for both the price and volume returns processes. Successively, we extend the mathematical model in a multivariate setting by considering a dependence structure between price, volumes and waiting times (durations) using a copula function.  

\subsection{Weighted-Indexed Semi-Markov Chains}

\indent Here, we introduce discrete-time WISMC model with Borel phase space in relation to the financial problem to which we are interested in.\\
\indent Let $(\Omega, \mathcal{F},\mathbb{P})$ be a probability space endowed with a filtration $\mathbb{F}:=(\mathcal{F}_{n})_{n\in \N}$ where all upcoming random variables are defined.\\ 
\indent Let $S(t)$ be the price of a financial asset at time $t\in \N$. The time varying log return, defined as $\log(S(t)/S(t-1))$, is usually the main variable object of investigation in financial literature. As commented in the previous section, at the short-time scales considered in high-frequency finance, this variable changes values only in correspondence of an increasing sequence of times $\{T_{n}^{J}\}_{n\in \N}$, the so-called jump-times of the asset price process.\\
\indent In correspondence of the times $\{T_{n}^{J}\}_{n\in \N}$, the logarithmic return process assumes different values denoted by $\{J_{n}\}_{n\in \N}$ and along any waiting time $X_{n}:=T_{n+1}^{J}-T_{n}^{J}$ it does not change value and remains constant. Thus, $J_{n}$ is the value of the logarithmic change in price at its n-th transition.\\
Let assume that at current time, say $t_{0}=0$, we dispose of a set of past data consisting of two vectors of observations collecting the last $m+1$ visited states of the log-return process and corresponding transitions times, respectively, i.e. 
\begin{equation*}
{\bf{J}}_{-m}^{0}=(J_{-m},J_{-m+1},\ldots, J_{0}),
\end{equation*}
\begin{equation*}
{\bf{T}}_{-m}^{0}=(T_{-m}^{J},T_{-m+1}^{J},\ldots, T_{0}^{J}).
\end{equation*}
Consider also an index process:
\begin{equation}
\label{funcrela}
I_{n}^{J}(\lambda):=\sum_{r=0}^{m+n-1}\sum_{a=T_{n-1-r}^{J}}^{T_{n-r}^{J}-1}f^{\lambda}(J_{n-1-r},T_{n}^{J},a)+f^{\lambda}(J_{n},T_{n}^{J},T_{n}^{J}),
\end{equation}
where $f^{\lambda}: \R \times \N \times \N \rightarrow \R$ is a bounded function.\\
\indent The process $I_{n}^{J}(\lambda)$ can be interpreted as an accumulated reward process with the function $f^{\lambda}$ as a measure of the weighted rate of reward per unit time. 
The parameter $\lambda$ is a memory parameter that should be calibrated on the data. A specific calibration procedure will be discussed in the application (Section 4). It should also be remarked that the index process considered in this paper is slightly more general than those considered in previous research articles because we added the term $f^{\lambda}(J_{n},T_{n}^{J},T_{n}^{J})$ that add to the index process also the score deriving from observing the current log-return state $J_{n}$ at present time $T_{n}^{J}$.\\
\indent Introduce the counting process $N^{J}(t):=\max\{n\in \N: T_{n}^{J}\leq t\}$, and let us now introduce the notion of weighted-indexed semi-Markov chains. 
\begin{definition} 
\label{wismc}
The process $Z^{J}(t):=J_{N^{J}(t)}$ is said to be a weighted-indexed semi-Markov chain with phase-space $(\R, \mathcal{B}(\R))$ if $\forall i,x,j\in \R$ and $\forall t\in \N$ there exists a function ${\bf{q}}^{J}=q^{J}(i,x;j,t)$, called the indexed semi-Markov kernel, such that $\forall n\in \N$ the following equality holds true:
\begin{equation}
\label{kernel}
\begin{aligned}
& \mathbb{P}[J_{n+1}\leq j,\: T_{n+1}^{J}-T_{n}^{J}= t |\sigma(J_{h},T_{h}^{J},I_{h}^{J}(\lambda), h\leq n), J_{n}=i,I_{n}^{J}(\lambda)=x]\\
& =\mathbb{P}[J_{n+1}\leq j,\: T_{n+1}^{J}-T_{n}^{J} = t |J_{n}=i,I_{n}^{J}(\lambda)=x]=:q^{J}(i,x;j,t).
\end{aligned}
\end{equation}
\end{definition}
\begin{remark}
\indent Relation $(\ref{kernel})$ asserts that the knowledge of the values of the variables $J_{n}, I_{n}^{J}(\lambda)$ is sufficient to give the conditional distribution of the couple $J_{n+1}, T_{n+1}^{J} - T_{n}^{J}$ whatever the values of the past variables might be. Therefore, to assess the probability of the next value of the log-return process and of the time in which the process is going to change state, we need only the knowledge of the last state of the log-return and the last value of the index process. 
\end{remark}
\begin{remark}
The function $Q^{J}(i,x;j,t):=\sum_{s\leq t}q^{J}(i,x;j,s)$ satisfies the following properties:\\
a) $Q^{J}(i,x;j,\cdot)$ is a nondecreasing discrete real function such that $$Q^{J}(i,x;j,0)=0.$$
b) $p^{J}(\cdot,x;\cdot):=Q^{J}(\cdot,x;\cdot,\infty)$ is a Markov transition probability function from $(\R, \mathcal{B}(\R))$ to itself.
\end{remark}
\begin{remark}
If the indexed semi-Markov kernel is constant in $x$, i.e. fixed the triple $(i,j,t)$ for all $y\neq x$ 
$$
q^{J}(i,x;j,t) \neq q^{J}(i,y;j,t),
$$
then, it degenerates in a semi-Markov kernel and the WISMC model becomes equivalent to classical semi-Markov chain model, see e.g. \cite{limn01} and \cite{dami17}.
\end{remark}
The triplet $\{J_{n}, T_{n}^{J}, I_{n}^{J}(\lambda)\}$ describes the system in correspondence of any jump time $T_{n}^{J}$. However, it is also important to describe the system in correspondence of any time $t$, which can be a jump time ($t = T_{n}^{J}$) or not ($t\neq T_{n}^{J}$). The random process $Z^{J}(t):=J_{N^{J}(t)}$ introduced in definition (\ref{wismc}) marks the log-return at any time $t$, while the backward recurrence time process $B^{J}(t):=t-T_{N^{J}(t)}^{J}$ denotes the time elapsed since the last transition. In our model this information is not sufficient to completely characterize the status of the system because we need to know also the value of the index process. To this end we extended the definition of the index process allowing to consider any time $t\in \N$ as follows:
\begin{equation}
\label{stocproc}
I^{J}(\lambda; t)=\sum_{r=0}^{m+N^{J}(t)-1+\theta}\,\,\sum_{a=T_{N^{J}(t)+\theta -1-r}^{J}}^{(t\wedge T_{N^{J}(t)+\theta-r}^{J})-1}f^{\lambda}(J_{N^{J}(t)+\theta-1-r},t,a)+f^{\lambda}(J_{N^{J}(t)},t,t),
\end{equation}
where $\theta =1_{\{t>T_{N^{J}(t)}^{J}\}}$. If $t=T_{n}^{J}$, that is $t$ is a jump time, we have that $I^{J}(\lambda; t)=I_{n}^{J}(\lambda)$.\\  
\indent The following definition and result, which reduces the complexity of the model, is important for practical application of the WISMC model.
\begin{definition}
\label{shift}[Shift operator]
Let $(i,t)_{-m}^{n}=\{(i_{\alpha},t_{\alpha}), \alpha =-m,\ldots,n\}$ be a sequence of states and corresponding transition times, i.e. $i_{\alpha}\in \R$, $t_{\alpha}\in \mathbb{Z}$, $t_{\alpha}<t_{\alpha +1}$. \\
Let denote by $\Theta_{-m}^{n}=\{(i_{\alpha}, t_{\alpha}), \alpha =-m,\ldots, 0,\ldots,n, i_{\alpha}\in \R, t_{\alpha}\in \mathbb{Z}\}$, then we define the shift operator
\begin{equation*}
\circ: \Theta_{-m}^{n+1}\rightarrow \Theta_{-m-1}^{n}
\end{equation*} 
defined by
\begin{equation*}
\circ((i,t)_{-m}^{n+1})=(s,k)_{-m-1}^{n}
\end{equation*}
where $s_{\alpha}=i_{\alpha +1}$, $k_{\alpha}=t_{\alpha +1}-t_{n+1}$,\,\,\,\, $\alpha =-m-1,\ldots,0,\ldots,n$.
\end{definition}
From an intuitive point of view, the shift operator when applied to a trajectory $(i,t)_{-m}^{n+1}$ gives back a new trajectory where the sequence of visited states is the same as in the input trajectory with the difference that transition times are translated of $t_{n+1}$ time units backward and the number of transitions is set one unit backward.\\
\indent The following assumption concerning the score function $f^{\lambda}$ will be needed in the rest of the article:\\
{\bf{A1}}: $\forall i\in \R, t\in \N, a\in \N$, $f^{\lambda}(i,t,a)=f^{\lambda}(i,t-a)$. 
\begin{lemma}
\label{homo}
For a WISMC with score function $f^{\lambda}$ that satisfies assumption {\bf{A1}}, for fixed arbitrary state $j$ and time $t$ and $(i,t)_{-m}^{n+1}\in \Theta_{-m}^{n+1}$, we have:  
\begin{equation}
\label{pr1}
\begin{aligned}
& \mathbb{P}[J_{n+2}\leq j,\: T_{n+2}^{J}-T_{n+1}^{J}= t |(J,T^{J})_{-m}^{n+1}=(i,t)_{-m}^{n+1}]\\
& =\mathbb{P}[J_{n+1}\leq j,\: T_{n+1}^{J}-T_{n}^{J}= t |(J,T^{J})_{-m-1}^{n}=\circ((i,t)_{-m}^{n+1})].
\end{aligned}
\end{equation}
\end{lemma}
\begin{proof}
See the appendix.
\end{proof}
The result presented in Lemma (\ref{homo}) focuses on a class of score functions leading probability $(\ref{pr1})$ to be independent of $n$. Accordingly, the WISMC inherits a homogeneity property that is particularly useful for the applications of the model. Throughout this article, we are going to consider homogeneous WISMC only.\\
\indent In this research we consider also financial volume as one important variable worthwhile to be investigated. The WISMC model was also applied to the modeling of financial volumes in a recent article by \cite{dape18b} and revealed to be able to reproduce several statistical properties of volumes at high-frequency scales. In order to be able to distinguish between the WISMC model for returns and that for volumes we introduce an additional notation for the volume model. Precisely,  if $V(t)$ is the volume of a financial asset at time $t\in \N$, the time varying log volume is defined as $\log(V(t)/V(t-1))$. This variable at short-time scales changes values in correspondence of an increasing sequence of times $\{T_{n}^{V}\}_{n\in \N}$, the so-called jump-times of the asset volume process. In correspondence of the times $\{T_{n}^{V}\}_{n\in \N}$, the logarithmic volume process assumes different values denoted by $\{J_{n}^{V}\}_{n\in \N}$. We introduce the index process for the volume by replacing in formula (\ref{funcrela}) the variables $J_{n}$, $T_{n}^{J}$ and $\lambda$ by $V_{n}$, $T_{n}^{V}$ and $\gamma$, respectively. The semi-Markov kernel for the volume process will be denotes by ${\bf{q}}^{V}=q^{V}(i,x;j,t)$ and the WISMC process for the volume variable is defined by $Z^{V}(t):=V_{N^{V}(t)}$ being $N^{V}(t):=\max\{n\in \N: T_{n}^{V}\leq t\}$.

\subsection{The multivariate model}\label{volume}

In this section we extend the WISMC model into a multivariate setting in such a way that it is able to describe jointly the time evolution of the three considered variables: log-returns, log-volumes and waiting times. The extension is done advancing a series of assumption that allow us to merge the WISMC kernel of the log-return process and that of the log-volumes in a new kernel that is completely characterized in this section.\\ 
\indent The first step in the joint modelization of returns, volumes and durations is to synchronize the time events of the returns and volumes. In order to do it let us start from the two sequences
\begin{equation}
(J_{n}, T_{n}^{J})_{n\in \N},\,\,\,\,\, (V_{n}, T_{n}^{V})_{n\in \N}.
\end{equation}
They mark the values and points in time where log-returns and log-volumes change states, respectively. First, we define a new sequence of transition times:
\begin{equation}
\label{time}
\{\tilde{T}_{n}\}=\{T_{n}^{J}\}\cup \{T_{n}^{V}\},\,\, \text{with}\,\, \tilde{T}_{0}=T_{0}^{J}=T_{0}^{V}=0.
\end{equation}
Relation (\ref{time}) means that we consider the union between the sets of transition times of returns and volumes and the obtained ordered sequence of times is denoted with the symbol $\{\tilde{T}_{n}\}_{n\in \N}$.  Intuitively, the time $\tilde{T}_{1}$ is the first time when a change in the returns or in the volumes occurred, $\tilde{T}_{2}$ the second point in time when a second change of state of whichever of the two processes $J_{n}$ and $V_{n}$ occurred, and so on. The corresponding inter-arrival times can be denoted by
\begin{equation*}
\tilde{X}_{n}=\tilde{T}_{n+1}-\tilde{T}_{n}.
\end{equation*}
\\
\indent Furthermore we define the corresponding values of the returns and volumes for each time of the random sequence $\tilde{T}_{n}$ according to the following relations: 
\begin{equation*}
\tilde{J}_{n}=J_{s},\,\, \text{if}\,\, s=\max \{h\in \N : T_{h}^{J}\leq \tilde{T}_{n}\},
\end{equation*}
\begin{equation*}
\tilde{V}_{n}=V_{s},\,\, \text{if}\,\, s=\max \{h\in \N : T_{h}^{V}\leq \tilde{T}_{n}\},
\end{equation*}
\indent Thus, we ended up with three variables $(\tilde{J}_{n},\tilde{V}_{n},\tilde{T}_{n})$ that denote the synchronized sequences of log-returns, log-volumes and transition times. In order to advance a joint model for this three-variate process we need to advance some specific properties concerning their interdependence and dynamics.\\
\indent Suppose the following conditional independence relation, namely assumption {\bf{A2}}, holds true: \\
\begin{equation}
\label{a22}
\begin{aligned}
&\mathbb{P}[\tilde{J}_{n+1}\leq j,\: \tilde{V}_{n+1}\leq a,\: \tilde{X}_{n}= t |\sigma(\tilde{J}_{h},\tilde{V}_{h},\tilde{T}_{h}, h\leq n), \tilde{J}_{n}=i, \tilde{V}_{n}=v,\tilde{I}_{n}^{J}=x, \\
& \tilde{I}_{n}^{V}=w, \tilde{T}_{n}=s, \tilde{T}_{n}-T_{N^{J}(s)}^{J}=b^{J}, \tilde{T}_{n}-T_{N^{V}(s)}^{V}=b^{V}]\\
& =\mathbb{P}[\tilde{J}_{n+1}\leq j,\: \tilde{V}_{n+1}\leq a,\: \tilde{X}_{n}= t  |\tilde{J}_{n}=i, \tilde{V}_{n}=v, \tilde{I}_{n}^{J}=x, \tilde{I}_{n}^{V}=w, \tilde{T}_{n}=s,\\
&  \tilde{T}_{n}-T_{N^{J}(s)}^{J}=b^{J}, \tilde{T}_{n}-T_{N^{V}(s)}^{V}=b^{V}].
\end{aligned}
\end{equation}
\indent Assumption {\bf{A2}} considers a quasi-Markovian-type hypothesis which asserts that the knowledge of the last values of synchronized variables $(\tilde{J}_{n}=i, \tilde{V}_{n}=v, \tilde{T}_{n}=s)$ together with corresponding values of the index processes $(\tilde{I}_{n}^{J}=x, \tilde{I}_{n}^{V}=w)$ and of the time elapsed from last transition of both log-return and log-volume $(\tilde{T}_{n}-T_{N^{J}(s)}^{J}=b^{J}, \tilde{T}_{n}-T_{N^{V}(s)}^{V}=b^{V})$ suffices to give the conditional distribution of the triplet $(\tilde{J}_{n+1},\: \tilde{V}_{n+1},\: \tilde{X}_{n})$ whatever the values of the past variables might be.\\
\indent The following information sets are introduced for notational convenience:
\[
\mathcal{A}_{n,s}^{J}:=\{\tilde{J}_{n}=i, \tilde{I}_{n}^{J}=x, \tilde{T}_{n}=s, \tilde{T}_{n}-T_{N^{J}(s)}^{J}=b^{J}\},
\]
\[
\mathcal{A}_{n,s}^{V}:=\{\tilde{V}_{n}=v, \tilde{I}_{n}^{V}=w, \tilde{T}_{n}=s, \tilde{T}_{n}-T_{N^{V}(s)}^{V}=b^{V}\},
\]
\[
\mathcal{A}_{n,s}^{JV}:=\mathcal{A}_{n,s}^{J} \bigcup \mathcal{A}_{n,s}^{V},
\]
\[
\mathcal{A}_{n,s}^{JVT}:=\mathcal{A}_{n,s}^{JV} \bigcup \{\tilde{X}_{n}=t\}.
\]
\indent Probability $(\ref{a22})$ is so important to merit a formal definition:
\begin{definition}
Let $(\tilde{J}_{n},\tilde{V}_{n},\tilde{T}_{n})$ be the synchronized triplet process of log-return, log-volume and transition times. The function 
\[
{\bf q}^{JV}=q^{JV}(\mathcal{A}_{n,s}^{JV};j,a,t),
\]
with $i,v,x,w,j,a\in \mathbb{R}$ and $b^{J},b^{V},t\in \mathbb{N}$ defined by
\begin{equation}
\label{kerdef}
\begin{aligned}
&  q^{JV}(\mathcal{A}_{n,s}^{JV};j,a,t):=\mathbb{P}[\tilde{J}_{n+1}\leq j,\: \tilde{V}_{n+1}\leq a,\: \tilde{X}_{n}= t  |\mathcal{A}_{n,s}^{JV}].
\end{aligned}
\end{equation}
is called the kernel of the triplet process. 
\end{definition}
\indent The kernel of the triplet process, can be factorized into the product of the conditional joint distribution of log-return and log-volumes multiplied by the conditional distribution of inter-arrival times, i.e.
\begin{equation}
\label{otto}
\mathbb{P}[\tilde{J}_{n+1}\leq j,\: \tilde{V}_{n+1}\leq a | \mathcal{A}_{n,s}^{JVT}]\cdot {\mathbb{P}[\tilde{X}_{n}= t |\mathcal{A}_{n,s}^{JV}]}.
\end{equation}
\indent Our next main task is to give a representation of this kernel in such a way that the dynamic of the joint process $(\tilde{J}_{n}, \tilde{V}_{n}, \tilde{T}_{n})$ could be completely characterized. We shall now consider reasonable data-driven assumptions that permits this computation.\\
\indent Assumption ${\bf{A3}}$: synchronized waiting time distribution.\\
\indent We assume that the probability distribution of inter-arrival time is independent on current time $s$, this avoid the use of time non-homogeneous probabilistic structures of the process. Moreover we assume that the waiting time distribution does not explicitly depends on the time elapsed by log-return and log-volume into their current states but includes past information depending on the index processes of returns and volumes. In formula 
\begin{equation}
{\mathbb{P}[\tilde{X}_{n}= t | \mathcal{A}_{n,s}^{JV}]}={\mathbb{P}[\tilde{X}_{n}= t | \tilde{J}_{n}=i, \tilde{V}_{n}=v, \tilde{I}_{n}^{J}=x, \tilde{I}_{n}^{V}=w]}.
\end{equation}
Assumption ${\bf{A3}}$ implies that the distributional properties of the waiting-times in our model can differ according to price and volume movements ($\tilde{J}_{n}$ and $\tilde{V}_{n}$ values) as well as with their past behavior measured by the index processes $\tilde{I}_{n}^{J}$ and $\tilde{I}_{n}^{V}$.\\
\indent Denote the conditional probability of $\tilde{X}_{n}$ by 
\begin{equation}
\label{acca}
\tilde{H}_{i,v}(x,w;t):={\mathbb{P}[\tilde{X}_{n}\leq t | \tilde{J}_{n}=i, \tilde{V}_{n}=v, \tilde{I}_{n}^{J}=x, \tilde{I}_{n}^{V}=w]},
\end{equation} 
 and the corresponding probability mass function by
\begin{equation}
\label{diffH}
\begin{aligned}
&{\mathbb{P}[\tilde{X}_{n}= t | \tilde{J}_{n}=i, \tilde{V}_{n}=v,\tilde{T}_{n}=s, \tilde{I}_{n}^{J}=x, \tilde{I}_{n}^{V}=w]}\\
& ={\tilde{H}_{i,v}(x,w;t)-\tilde{H}_{i,v}(x,w;t-1)}=:\tilde{h}_{i,v}(x,w;t).
\end{aligned}
\end{equation}
\indent The independence of the cdf of waiting times on the number of transitions $n$, and on the time of last transition $s$ is done in order to avoid unnecessary complications that would have made the model inhomogeneous in time.\\
\indent The knowledge of the kernel of the triplet process needs also the specification of the conditional joint probability distribution of log-returns and log-volumes. In this respect we propose the following \\
\indent Assumption ${\bf{A4}}$: the conditional joint distribution of modulus of log-returns and log-volumes is given by
\begin{equation}
\label{margcop}
\begin{aligned}
\mathbb{P}[\mid \tilde{J}_{n+1}\mid \leq j,\: \mid \tilde{V}_{n+1}\mid \leq a\, |\, \mathcal{A}_{n,s}^{JVT}]=\mathcal{C}\Big(F^{|J|}(i,x,t+b^{J};j), F^{|V|}(v,w,t+b^{V};a\Big),
\end{aligned}
\end{equation}
where $\mathcal{C}$ is a Copula-function and the marginal distributions $F^{|J|}(i,x,t+b^{J};j)$ and $F^{|V|}(v,w,t+b^{V};a)$ are given by  
\begin{equation*}
\begin{aligned}
& F^{|J|} = \mathbb{P}[\mid {J}_{N^{J}(s)+1}\mid \leq j \, |\, T_{N^{J}(s)+1}-s= t, {J}_{N^{J}(s)}=i, {I}_{N^{J}(s)}^{J}=x, s-T_{N^{J}(s)}^{J}=b^{J}],\\
& F^{|V|} = \mathbb{P}[\mid {V}_{N^{V}(s)+1}\mid \leq a \, |\, T_{N^{V}(s)+1}-s= t, {V}_{N^{V}(s)}=i, {I}_{N^{V}(s)}^{V}=x,  s-T_{N^{V}(s)}^{V}=b^{V}].
\end{aligned}
\end{equation*}
\indent Assumption ${\bf{A4}}$ is motivated by the data analysis executed in Section 2, specifically in Table \ref{table_cross_corr} we have shown that the two processes are dependent on each other. Essentially this assumption allows us to consider a dependence structure between the modulus of the log-returns and log-volumes that is managed through the use of any copula function. The copula maps the two marginal distributions $F^{|J|}$ and $F^{|V|}$ into a joint probability distribution function.  The quantity $F^{|J|}$ expresses the probability to get the modulus of log-return less or equal to $j$ conditionally on the last value of the variable, corresponding index process, waiting time length and duration in the last visited states. The same interpretation can be given to the quantity $F^{|V|}$ with the only exception that it is related to the modulus of log-volume process.\\
\indent The $F^{|J|}$ can be evaluated as follows:
\begin{equation*}
\begin{aligned}
& F^{|J|}(i,x,t+b^{J};j)\\
& = \mathbb{P}[\mid {J}_{N^{J}(s)+1}\mid \leq j \, |\, T_{N^{J}(s)+1}-s= t, {J}_{N^{J}(s)}=i, {I}_{N^{J}(s)}^{J}=x, s-T_{N^{J}(s)}^{J}=b^{J}]\\
& =\mathbb{P}[\mid {J}_{N^{J}(s)+1}\mid \leq j \, |\, T_{N^{J}(s)+1}-T_{N^{J}(s)}+T_{N^{J}(s)}-s= t, {J}_{N^{J}(s)}=i,\\
&  {I}_{N^{J}(s)}^{J}=x, T_{N^{J}(s)}^{J}=s-b^{J}]\\
& = \mathbb{P}[\mid {J}_{N^{J}(s)+1}\mid \leq j \, |\, T_{N^{J}(s)+1}-T_{N^{J}(s)}= t+b^{J}, {J}_{N^{J}(s)}=i, {I}_{N^{J}(s)}^{J}=x]\\
& = \mathbb{P}[\mid {J}_{n+1}\mid \leq j \, |\, T_{n+1}-T_{n}= t+b^{J}, {J}_{n}=i, {I}_{n}^{J}=x]\\
& = \frac{\mathbb{P}[\mid {J}_{n+1}\mid \leq j, T_{n+1}-T_{n}= t+b^{J} \, |\,  {J}_{n}=i, {I}_{n}^{J}=x]}{\mathbb{P}[T_{n+1}-T_{n}= t+b^{J} \, |\,  {J}_{n}=i, {I}_{n}^{J}=x]}\\
& = \frac{q^{J}(i,x;j,t+b^{J})-q^{J}(i,x;-j,t+b^{J})}{H_{i}^{J}(x; t+b^{J})-H_{i}^{J}(x; t+b^{J}-1)}.
\end{aligned}
\end{equation*}
\indent Similar computations gives
\begin{equation*}
\begin{aligned}
F^{|V|}(v,w,t+b^{V};a)=\frac{q^{V}(v,w;a,t+b^{V})-q^{V}(v,w;-a,t+b^{V})}{H_{v}^{V}(w; t+b^{V})-H_{v}^{V}(w; t+b^{V}-1)}.
\end{aligned}
\end{equation*}
\indent By means of assumptions {\bf{A3}} and {\bf{A4}} we can get information on the joint distribution of modulus of log-returns and modulus of log-volumes. Nonetheless, it is our interest to recover information on the exact values (with signs) of these two variables. This is motivated by the empirical observation that although $\{\mid \tilde{J}_{n}\mid \}$ and $\{\mid \tilde{V}_{n}\mid \}$ are significantly correlated, $\{\tilde{J}_{n}\}$ and $\{\tilde{V}_{n}\}$ are uncorrelated.\\
\indent To be able to reach this objective we advance a final assumptions:\\
 ${\bf{A5}}$: For each $n \in \mathbb{N}$, $\tilde{J}_{n}$ and $\tilde{V}_{n}$ satisfy the following relations:
\begin{equation*}
\begin{aligned}
& \tilde{J}_{n}=\mid \tilde{J}_{n} \mid \cdot \, \eta_{n}^{J};\\
& \tilde{V}_{n}=\mid \tilde{V}_{n} \mid \cdot \, \eta_{n}^{V};
\end{aligned}
\end{equation*}
where $\eta_{n}^{J}$ and $\eta_{n}^{V}$ are two sequences of i.i.d. random variables with pmf
\begin{eqnarray*}
&&\eta_{n}^{J}\sim \left\{
                \begin{array}{cl}
                       \ +1  &\mbox{with probability $p^{J}$}\\
                         -1  &\mbox{with probability $1-p^{J}$}\\
                   \end{array}
             \right.
\end{eqnarray*}
\begin{eqnarray*}
&&\eta_{n}^{V}\sim \left\{
                \begin{array}{cl}
                       \ +1  &\mbox{with probability $p^{V}$}\\
                         -1  &\mbox{with probability $1-p^{V}$}\\
                   \end{array}
             \right.
\end{eqnarray*}
\indent This assumptions allows us to get the value of the variables starting from the knowledge of their modulus. Indeed, the variables $\eta_{n}^{J}$ and $\eta_{n}^{V}$ provides the sign of the size of the variation. Obviously, the parameters $p^{J}$ and $p^{V}$ need to be estimated on the data.\\  
\indent The next theorem will characterize the kernel of the triplet process.
\begin{theorem}
\label{kernelproof}
Under assumptions ${\bf{A1-A5}}$, $\forall s\in \N$ the kernel of the triplet process $(\tilde{J}_{n},\tilde{V}_{n},\tilde{T}_{n})$, defined in formula $(\ref{kerdef})$, is given by:
\begin{equation}
\label{kerntripletI}
\begin{aligned}
& \text{(i)}\,\, \text{for} \,\, j\geq 0, a\geq 0\\
&q^{JV}(\mathcal{A}_{n,s}^{JV};j,a,t) = \tilde{h}_{i,v}(x,w;t) \cdot \Bigg[ 1+ p^{J}p^{V} \Bigg(1-F^{|J|}(i,x,t+b^{J};j)\\
& -F^{|V|}(v,w,t+b^{V};a)+\mathcal{C}\Big(F^{|J|}(i,x,t+b^{J};j), F^{|V|}(v,w,t+b^{V};a)\Big)\Bigg)\\
& -p^{V}\Big(1-F^{|V|}(v,w,t+b^{V};a)\Big)-p^{J}\Big(1-F^{|J|}(i,x,t+b^{J};j)\Big)\Bigg],
\end{aligned}
\end{equation}
\begin{equation}
\label{kerntripletII}
\begin{aligned}
& \text{(ii)}\,\, \text{for} \,\, j< 0, a< 0\\
&q^{JV}(\mathcal{A}_{n,s}^{JV};j,a,t)=\tilde{h}_{i,v}(x,w;t) \cdot (1- p^{J})(1- p^{V}) \Bigg[ 1-F^{|J|}(i,x,t+b^{J};-j)\\
& -F^{|V|}(v,w,t+b^{V};-a)+\mathcal{C}\Big(F^{|J|}(i,x,t+b^{J};-j), F^{|V|}(v,w,t+b^{V};-a)\Big)\Bigg],
\end{aligned}
\end{equation}
\begin{equation}
\label{kerntripletIII}
\begin{aligned}
& \text{(iii)}\,\, \text{for} \,\, j< 0, a> 0\\
&q^{JV}(\mathcal{A}_{n,s}^{JV};j,a,t)= \tilde{h}_{i,v}(x,w;t) \cdot \Bigg[(1- p^{J})\cdot \Big[F^{|V|}(v,w,t+b^{V};a)\\
& -\mathcal{C}\Big(F^{|J|}(i,x,t+b^{J};-j), F^{|V|}(v,w,t+b^{V};a)\Big)\Big]+ (1- p^{J})(1- p^{V})\\
& \cdot \Big[ 1-F^{|J|}(i,x,t+b^{J};-j) -F^{|V|}(v,w,t+b^{V};a)\\
& +\mathcal{C}\Big(F^{|J|}(i,x,t+b^{J};-j), F^{|V|}(v,w,t+b^{V};a)\Big)\Big]\Bigg],
\end{aligned}
\end{equation}
\begin{equation}
\label{kerntripletIV}
\begin{aligned}
& \text{(iv)}\,\, \text{for} \,\, j> 0, a< 0\\
&q^{JV}(\mathcal{A}_{n,s}^{JV};j,a,t)= \tilde{h}_{i,v}(x,w;t) \cdot \Bigg[(1- p^{V})\cdot \Big[F^{|J|}(i,x,t+b^{J};j)\\
& -\mathcal{C}\Big(F^{|J|}(i,x,t+b^{J};j), F^{|V|}(v,w,t+b^{V};-a)\Big)\Big]+ (1- p^{J})(1- p^{V})\\
& \cdot \Big[ 1-F^{|J|}(i,x,t+b^{J};j) -F^{|V|}(v,w,t+b^{V};-a)\\
& +\mathcal{C}\Big(F^{|J|}(i,x,t+b^{J};j), F^{|V|}(v,w,t+b^{V};-a)\Big)\Big]\Bigg],
\end{aligned}
\end{equation}
\end{theorem}
\begin{proof}
See the appendix.
\end{proof}

\subsection{Financial functions}
In this subsection we show how it is possible to compute financial functions of specific interest using the characterization of the kernel of the triplet process given in Theorem \ref{kernelproof}. Results are confined to marginal distributions of log-returns and log-volumes, correlation structures and joint first passage time distributions. In general given the kernel, it is possible to compute any type of functional of the kernel.

\subsubsection{The one-step marginal distributions of returns and volumes}
The first question to which we are interested in is the determination of the marginal distributions of log-returns and log-volumes. Since the dependence structure has been introduced on the modulus of these variables the marginal distributions we are looking for do not coincide with those used in the copula, i.e. with $F^{|J|}$ and $F^{|V|}$.\\
\indent Let us consider the problem of finding the marginal distribution of the return process. Let us proceed by integration of the volume variable and summation on the duration one, this gives $\sum_{t\geq 0}q^{JV}(\mathcal{A}_{n,s}^{JV};j,\infty,t)$. Thus, for $j>0$, using the kernel representation $(\ref{kerntripletI})$ and the fact that $F^{|V|}(v,w,|\infty|,t+b^{V})=1$ and that $\mathcal{C}\Big(F^{|J|}(i,x,t+b^{J};j), F^{|V|}(v,w,t+b^{V};\infty)\Big)=F^{|J|}(i,x,t+b^{J};j)$ we obtain the following sequence of equalities:
\begin{equation*}
\begin{aligned}
& \sum_{t\geq 0}q^{JV}(\mathcal{A}_{n,s}^{JV};j,\infty,t)=\sum_{t\geq 0}\tilde{h}_{i,v}(x,w;t) \cdot \Bigg[ 1+ p^{J}p^{V} \Bigg(1-F^{|J|}(i,x,t+b^{J};j)\\
& -1+F^{|J|}(i,x,t+b^{J};j)\Bigg) -p^{V}\Big(1-1)\Big)-p^{J}\Big(1-F^{|J|}(i,x,t+b^{J};j)\Big)\Bigg]\\
&=\sum_{t\geq 0} \tilde{h}_{i,v}(x,w;t) \cdot \Bigg[ 1 - p^{J}\Big(1-F^{|J|}(i,x,t+b^{J};j)\Big)\Bigg]\\
&=1- p^{J}\cdot \Bigg(1-\sum_{t\geq 0} \tilde{h}_{i,v}(x,w;t) \cdot F^{|J|}(i,x,t+b^{J};j)\Bigg).
\end{aligned}
\end{equation*} 
\indent This marginal distribution expresses the probability to observe with next transition, executed at any future time $t$, a return not greater than $j$. Symmetric arguments can be used to get the marginal distrbution of volumes that results in
\[
1- p^{V}\cdot \Bigg(1-\sum_{t\geq 0} \tilde{h}_{i,v}(x,w;t) \cdot F^{|V|}(v,w,t+b^{V};a)\Bigg).
\]

\subsubsection{Dependence measures}
The kernel of the triplet process $(\ref{kerdef})$ completely describes the dependence structure between returns and volumes and waiting times. Nevertheless, it is relevant to measure this dependence using classical indicators of linear and nonlinear dependence.\\
\indent The most widely studied measure of linear dependence is the correlation coefficient. Let $\rho_{\mathcal{A}_{n,s}^{JV}}(|\tilde{J}_{n+1}|,|\tilde{V}_{n+1}|)$ be the correlation coefficient between the modulus of returns and the modulus of volumes at next transition unconditionally on the time when the next transition will happen, i.e.
\begin{equation}
\rho_{\mathcal{A}_{n,s}^{JV}}(|\tilde{J}_{n+1}|,|\tilde{V}_{n+1}|)=\frac{Cov_{\mathcal{A}_{n,s}^{JV}}(|\tilde{J}_{n+1}|,|\tilde{V}_{n+1}|)}{\sigma_{\mathcal{A}_{n,s}^{J}}(|\tilde{J}_{n+1}|)\cdot \sigma_{\mathcal{A}_{n,s}^{V}}(|\tilde{V}_{n+1}|)}.
\end{equation}  
\indent Using the formula discussed above, we can calculate the correlation coefficient by using the joint probability density function of $(|\tilde{J}_{n+1}|,|\tilde{V}_{n+1}|)$ conditional on the information set $\mathcal{A}_{n,s}^{JV}$. For every $j,a\geq 0$, one gets:
\begin{equation}
\begin{aligned}
&F_{(|\tilde{J}_{n+1}|,|\tilde{V}_{n+1}|)}(j,a):=\mathbb{P}[\mid \tilde{J}_{n+1}\mid \leq j,\: \mid \tilde{V}_{n+1}\mid \leq a\, |\, \mathcal{A}_{n,s}^{JV}]\\
& \sum_{t\geq 0}\mathbb{P}[\mid \tilde{J}_{n+1}\mid \leq j,\: \mid \tilde{V}_{n+1}\mid \leq a\, |\, \mathcal{A}_{n,s}^{JVT}]\cdot \mathbb{P}[\tilde{X}_{n}= t\, |\, \mathcal{A}_{n,s}^{JV}]\\
& \sum_{t\geq 0} \tilde{h}_{i,v}(x,w;t) \cdot \mathcal{C}\Big(F^{|J|}(i,x,t+b^{J};j), F^{|V|}(v,w,t+b^{V};a\Big).
\end{aligned}
\end{equation}
\indent Consequently the density can be obtained by derivation of the cumulative distribution function, i.e.
\begin{equation}
\label{density}
\begin{aligned}
& f_{(|\tilde{J}_{n+1}|,|\tilde{V}_{n+1}|)}(j,a)=\frac{\partial^{2}}{\partial j \partial a}F_{(|\tilde{J}_{n+1}|,|\tilde{V}_{n+1}|)}(j,a)\\
& =\sum_{t\geq 0} \tilde{h}_{i,v}(x,w;t) \cdot \frac{\partial^{2}}{\partial j \partial a}\mathcal{C}\Big(F^{|J|}(i,x,t+b^{J};j), F^{|V|}(v,w,t+b^{V};a\Big).
\end{aligned}
\end{equation}
\indent Accordingly we get
\begin{equation}
\begin{aligned}
Cov_{\mathcal{A}_{n,s}^{JV}}(|\tilde{J}_{n+1}|,|\tilde{V}_{n+1}|)&=\int_{0}^{\infty}\int_{0}^{\infty}j\cdot a \cdot f_{(|\tilde{J}_{n+1}|,|\tilde{V}_{n+1}|)}(j,a) djda\\
& - \int_{0}^{\infty}j  f_{|\tilde{J}_{n+1}|}(j) dj \cdot \int_{0}^{\infty}a f_{|\tilde{V}_{n+1}|}(a) da .
\end{aligned}
\end{equation}
\indent This allows the recovering of $\rho_{\mathcal{A}_{n,s}^{JV}}(|\tilde{J}_{n+1}|,|\tilde{V}_{n+1}|)$ once the standard deviations $\sigma_{\mathcal{A}_{n,s}^{J}}(|\tilde{J}_{n+1}|)$ and $\sigma_{\mathcal{A}_{n,s}^{V}}(|\tilde{V}_{n+1}|)$ are known. They can be obtained by using the univariate densities $f_{|\tilde{J}_{n+1}|}(\cdot)$ and $f_{|\tilde{V}_{n+1}|}(\cdot)$ that in turn can be obtained by integration of the joint density.\\
\indent It is also interesting to compute the covariance function between the modulus of log-returns and the log-volumes at next transition, i.e.
\begin{equation}
\begin{aligned}
& Cov_{\mathcal{A}_{n,s}^{JV}}(|\tilde{J}_{n+1}|,\tilde{V}_{n+1})=\mathbb{E}_{\mathcal{A}_{n,s}^{JV}}[|\tilde{J}_{n+1}|\cdot \tilde{V}_{n+1}]-\mathbb{E}_{\mathcal{A}_{n,s}^{J}}[|\tilde{J}_{n+1}|]\cdot \mathbb{E}_{\mathcal{A}_{n,s}^{V}}[\tilde{V}_{n+1}]\\
& = \mathbb{E}_{\mathcal{A}_{n,s}^{JV}}[|\tilde{J}_{n+1}|\cdot \eta_{n+1}^{V}\cdot |\tilde{V}_{n+1}|]-\mathbb{E}_{\mathcal{A}_{n,s}^{J}}[|\tilde{J}_{n+1}|]\cdot \mathbb{E}_{\mathcal{A}_{n,s}^{V}}[\eta_{n+1}^{V}\cdot |\tilde{V}_{n+1}|]\\
& = \mathbb{E}_{\mathcal{A}_{n,s}^{V}}[\eta_{n+1}^{V}]\cdot \Big(Cov_{\mathcal{A}_{n,s}^{JV}}(|\tilde{J}_{n+1}|,|\tilde{V}_{n+1}|)\Big).
\end{aligned}
\end{equation}
\indent Note that if $\mathbb{E}_{\mathcal{A}_{n,s}^{V}}[\eta_{n+1}^{V}]=0$, then the modulus of log-returns and log-volumes are uncorrelated at next transition.\\
\indent One may also be interested in providing nonlinear measures of dependence between random variables. Mutual information, which goes back to \cite{shannon}, possesses relevant properties that imposed it as a suitable measure of nonlinear dependence, see e.g. \cite{dion04}. It is simple to express the mutual information within our model:
\begin{equation}
\begin{aligned}
& MI_{\mathcal{A}_{n,s}^{JV}}(|\tilde{J}_{n+1}|,|\tilde{V}_{n+1}|)=\int_{0}^{\infty}\int_{0}^{\infty}f_{(|\tilde{J}_{n+1}|,|\tilde{V}_{n+1}|)}(j,a) \log \frac{f_{(|\tilde{J}_{n+1}|,|\tilde{V}_{n+1}|)}(j,a)}{f_{|\tilde{J}_{n+1}|}(j) \cdot f_{|\tilde{V}_{n+1}|}(a)} djda ,
\end{aligned}
\end{equation}
\noindent where the densities are given in formula $(\ref{density})$.

\subsubsection{First passage time distributions}\label{sec_fptd}

The first passage time distribution has attracted a lot of attention in finance. It has been considered for different assumptions about the stochastic processes that describes the asset behaviour. It has been investigated for log-returns when described by Ornstein-Uhlenbeck processes (see e.g. \cite{yi10}) and more recently for generalized semi-Markov models in \cite{dape11,dape12,dape12b}. We shall now derive the first passage time distribution for our multivariate model.\\
\indent Let $\tilde{M}_{t}^{J}(\tau)$ be the accumulation factor of the return process in the multivariate model from time $t$ to $t+\tau$. Formally, the accumulation factor can be defined as follows:
\begin{equation*}
\tilde{M}_{t}^{J}(\tau) = e^{\sum_{r=0}^{\tau -1}\tilde{Z}^{J}(t+r)}.
\end{equation*}
\indent A similar definition applies for the volume process, i.e. 
\begin{equation*}
\tilde{M}_{t}^{V}(\tau) = e^{\sum_{r=0}^{\tau -1}\tilde{Z}^{V}(t+r)}.
\end{equation*}
\indent For $\rho \in \R_{+}$ and $\psi \in \R_{+}$, denote the joint first passage time by 
\begin{equation}
\label{fptbid}
\Gamma_{(\rho ; \psi)}:=\min\{\tau \geq 0 : \{\tilde{M}_{0}^{J}(\tau) \geq \rho\} \cup \{\tilde{M}_{0}^{V}(\tau) \geq \psi\}\}.
\end{equation}
\indent Thus, $\Gamma_{(\rho ; \psi)}$ is the first time when at least one accumulation factor exceeds its own thresholds. Denote the corresponding conditional survival function by
\begin{equation*}
R_{(\rho ; \psi)}((i,v,t)_{-m}^{0},u;t)=\mathbb{P}[\Gamma_{(\rho ; \psi)} >t|(\tilde{J},\tilde{V},\tilde{T})_{-m}^{0}=(i,v,t)_{-m}^{0},\tilde{B}(u)=u],
\end{equation*}
where $\tilde{B}(u)=u-\tilde{T}_{\tilde{N}(u)}$.\\
\indent The definition of the shift operator given in Definition \ref{shift} can be easily extended to triplet sequences $(i,v,t)_{-m}^{n}$.  
\begin{definition}
\label{sequence3}
Let $(i,v,t)_{-m}^{n}=\{(i_{\alpha},v_{\alpha}, t_{\alpha}), \alpha =-m,\ldots,n\}$ be a sequence of returns, volumes and corresponding transition times.\\
Let denote by $\Phi_{-m}^{n}=\{(i_{\alpha}, v_{\alpha}, t_{\alpha}), \alpha =-m,\ldots, 0,\ldots,n, i_{\alpha}\in \R, v_{\alpha}\in \R, t_{\alpha}\in \mathbb{Z}, t_{\alpha}<t_{\alpha +1}\}$, then we define the shift operator
\begin{equation*}
\circ: \Phi_{-m}^{n+1}\rightarrow \Phi_{-m-1}^{n}
\end{equation*} 
defined by
\begin{equation*}
\circ((i,v,t)_{-m}^{n+1})=(s,y,k)_{-m-1}^{n}
\end{equation*}
where $s_{\alpha}=i_{\alpha +1}$, $y_{\alpha}=v_{\alpha +1}$, $k_{\alpha}=t_{\alpha +1}-t_{n+1}$,\,\,\,\, $\alpha =-m-1,\ldots,0,\ldots,n$.
\end{definition}
We formulate and prove a theorem which provides an equation for the joint first passage time distribution. 
\begin{theorem}
\label{prop}
Let $f^{\lambda}$ and  $g^{\gamma}$ be the score functions of the index processes relative to the return and volume processes, respectively. For ${i_{0}\geq 0}$ and ${v_{0}\geq 0}$, it results that
\begin{equation}
\label{fptdistri}
\begin{aligned}
& R_{(\rho ; \psi)}((i,v,t)_{-m}^{0},u;t) ={1_{\{e^{i_{0}t}<\rho\}}}{1_{\{e^{v_{0}t}<\psi\}}}\frac{1-\tilde{H}_{i_{0},v_{0}}\big(\alpha_{0},\beta_{0}; t\big)}{1-\tilde{H}_{i_{0},v_{0}}\big(\alpha_{0},\beta_{0}; u\big)}\\
& +\sum_{t_{1}=u+1}^{t}\int_{-\infty}^{+\infty}\int_{-\infty}^{+\infty}di_{1}dv_{1}
\frac{{1_{\{e^{i_{0}t_{1}}<\rho\}}}{1_{\{e^{v_{0}t_{1}}<\psi\}}}}{1-\tilde{H}_{i_{0},v_{0}}\big(\alpha_{0},\beta_{0}; u\big)}\\
& \cdot \frac{\partial^{2}q^{JV}(i_{0},v_{0},\alpha_{0},\beta_{0};i_{1},v_{1},t_{1})}{\partial i_{1}\partial v_{1}} \cdot R_{\Big(\frac{\rho}{e^{i_{0}t_{1}}};\frac{\psi}{e^{v_{0}t_{1}}}\Big)}(\circ((i,v,t)_{-m}^{1}),0;t-t_{1}),
\end{aligned}
\end{equation}
where
\begin{equation}
\begin{aligned}
& \alpha_{0}=\sum_{r=0}^{m-1}\sum_{a=t_{-r-1}}^{t_{-r}-1}f^{\lambda}(i_{-r-1},-a)+f^{\lambda}(i_{0},0),\\
& \beta_{0}=\sum_{r=0}^{m-1}\sum_{a=t_{-r-1}}^{t_{-r}-1}g^{\gamma}(v_{-r-1},-a)+g^{\gamma}(v_{0},0).
\end{aligned}
\end{equation}
\indent For $i_{0}<0$ replace $e^{i_{0}t_{1}}$ with $e^{i_{0}}$ everywhere in formula $(\ref{fptdistri})$.\\
\indent For $v_{0}<0$ replace $e^{v_{0}t_{1}}$ with $e^{v_{0}}$ everywhere in formula $(\ref{fptdistri})$.\\
\end{theorem}
\begin{proof}
See the appendix.
\end{proof}

\section{Application to real high frequency data}

To verify the validity of the model described above, we applied it to the database introduced in Section \ref{database}. 
Following \cite{dape12b, dape18b} we use, as definition of the function $f^{\lambda}$ in  (\ref{funcrela}), an exponentially weighted moving average (EWMA) of the squares of $J_{n}$ which has the following expression:
\begin{equation}
\label{funct}
f^{\lambda}(J_{n-1-k},T_{n}^{J},a)=\frac{\lambda^{T_{n}^{J}-a} J_{n-1-k}^2}{\sum_{k=0}^{m+n-1}\sum_{a=T_{n-1-k}^{J}}^{T_{n-k}^{J}-1}\lambda^{T_{n}^{J}-a}+1}=\frac{\lambda^{T_{n}^{J}-a} J_{n-1-k}^2}{\sum_{a=T_{-m}^{J}}^{T_{n}^{J}}\lambda^{a}}.
\end{equation}

A similar choice is done for the volume return process leading to the choice of 

\begin{equation}
\label{functV}
g^{\gamma}(V_{n-1-k},T_{n}^{V},a)=\frac{\gamma^{T_{n}^{V}-a} V_{n-1-k}^2}{\sum_{k=0}^{m+n-1}\sum_{a=T_{n-1-k}^{V}}^{T_{n-k}^{V}-1}\gamma^{T_{n}^{V}-a}+1}=\frac{\gamma^{T_{n}^{V}-a} V_{n-1-k}^2}{\sum_{a=T_{-m}^{V}}^{T_{n}^{V}}\gamma^{a}}.
\end{equation}

We remark that the choice for the functional form of $f^{\lambda}$ is not obtained trough any optimization procedure.
One can probably find functional forms that perform better according to some performance measure. Our choice
is motivated by its simplicity and the fact that it gives very good results.
Moreover, it is justified by the empirical evidence that price and volumes returns dynamics do depend on volatility regime.


Next step in the application is finding the optimal parameters to be used in the model. We followed the same procedure described in \cite{dape_18} and summarized in the following subsection.

\subsection{Parameters optimization}
We describe here the whole procedure to set and optimize the parameters used in the univariate models. 
\begin{enumerate}
	\item The first step to set the WISMC model is, by using the descriptive statistics of the dataset, to fix a number of states $s$ and a value for the weight parameter $\lambda$;
	\item Build the trajectory $(J_{n}, T_{n}^J)$ implied by the choice of $s$ and $\lambda$;
	\item Estimate the weighted-indexed semi-Markov kernel ${\bf{q}}^{J}$ applying the empirical estimators to the trajectory obtained at previous step;
	\item Perform Monte Carlo simulation to build synthetic time series;
	\item Estimate the autocorrelation function (ACF) for the synthetic time series $\Sigma(\tau; s, \lambda)$. Note that this ACF depends on the number of states $s$ and on the value of the weight parameter $\lambda$;
	\item Compare the real ACF, $\Sigma(\tau)$, with the synthetic one, $\Sigma(\tau; s, \lambda)$, by computing the Mean Absolute Percentage Error (MAPE) between them. The MAPE depends on the number of states and on the value of the weight parameter, then it is denoted by $MAPE(s, \lambda)$;
	\item Change the number of states and the parameter $\lambda$, restart from point 2 and repeat all points;
\end{enumerate}
At the end of the whole process, choose the number of states $s^{*}$ and parameter $\lambda^{*}$ that best represent the dataset by minimizing the $MAPE(s,\lambda)$, i.e.
$$(s^{*},\lambda^{*})=\argmin_{(s,\lambda)} \{MAPE(s,\lambda)\}.$$
Notice that the algorithm can stop whenever the increase in the number of states does not decrease the MAPE more than a given threshold $\epsilon$.\\
This procedure should be repeated for all stocks in the portfolio and also for the variable $v(t)$. 
Once all the parameters for the two univariate models are optimized use a copula to build the bivariate model.

\subsection{Results}
Here we show some results obtained and a comparison with real data. 
Using the optimization procedure described above we found the optimal parameters which are summarized in table \ref{opt_par} for the four stocks

\begin{table}
	\begin{center}
		\scalebox{0.9}{
			\begin{tabular}{|l||l|l|l||l|l|l|}
				\hline
				&\multicolumn{3}{|c|}{$r(t)$}&\multicolumn{3}{|c|}{$v(t)$} \\
				\cline{2-7}
				&$s$&$\lambda$&$MAPE (\%)$&$s$&$\gamma$&$MAPE (\%)$\\
				\hline
				\textbf{TIT}&5&0.97&7.7&5&0.97&10.2\\\hline
				\textbf{ISP}&5&0.98&4.6&5&0.98&6.1\\\hline	
				\textbf{TEN}&5&0.97&6.2&7&0.97&9.2\\\hline	
				\textbf{F}&5&0.97&3.7&9&0.97&5.5\\\hline	
		\end{tabular}}
	\end{center}
	\caption{Parameters used in the application to real data.}\label{opt_par}
\end{table}

The dependence between the two real processes $v(t)$ and $r(t)$ is kept in the model by using a copula function. We tested different copulas like Gaussian, t-student, Gumbel and Clayton finding almost no differences in the results. This is mainly due to the fact that 1 minute price returns are almost discrete and varies in a small range, then, in this dataset there is no tail effect. To keep the application as simple as possible we decided to use a Gaussian copula that has only one parameter. We simulated, using the estimated kernels and a Gaussian copula, the joint process $|r(t)|$ and $v(t)$ and obtained $r(t)$ by using the relation described in Assumption A5.  The results are trajectories with the same time length of real data for both variables $v(t)$ and $r(t)$. 

In Table \ref{table_cross_corr_sim} we show the cross-correlation between the synthetic $v(t)$ and $r(t)$ and their absolute values, for all combinations, as done in Table \ref{table_cross_corr}. The Table shows that there is a good agreement with what was found for real data.

\begin{table}
	\begin{center}
		\scalebox{0.7}{
			\begin{tabular}{|l|l|l|l|l|l|l|l|l|}
				\hline
				&$\rho(r(t),v(t))$&p-value&$\rho(|r(t)|,v(t))$&p-value& $\rho(r(t),|v(t))|$&p-value&$\rho(|r(t)|,|v(t))|$&p-value\\\hline
				TIT&0.0001&0&0.091&0&-0.0042&0.36&0.0039&0\\\hline
				ISP&-0.00390&0&0.083&0&-0.0025&0.053&-0.0086&0\\\hline
				TEN&-0.0046&0&0.081&0&0.0011&0.46&0.022&0\\\hline
				F&-0.0075&0&0.12&0&-0.0029&0.025&-0.0092&0\\\hline
		\end{tabular}}
	\end{center}
	\caption{Cross correlation between price and volume returns for simulated data.}\label{table_cross_corr_sim}
\end{table}

The model is also used to compare the first passage time distribution (fptd) of the joint processes $v(t)$ and $r(t)$. 
From the synthetic variables ($r(t)$ and $v(t)$) we build the variables price and volumes in the following way: at each discrete state of $r(t)$ ($v(t)$) is associated a range of variability of the continuous real $r(t)$ ($v(t)$),  inside this range a continuous value is chosen by extracting a random number form a uniform distribution and then inverting the empirical distribution of real continuous price (volume) return. Once $r(t)$ ($v(t)$) are transformed back into continuous values, prices $S(t)$ (volume $V(t)$) are obtained by $S(t) = S_0\times e^{\sum_{k=1}^{k=t}r(k)}$ ($V(t) = V_0\times e^{\sum_{k=1}^{k=t}v(k)}$.). 
Synthetic price and volumes are then used to build the distribution of time at which there is a first cross of given thresholds. 

\begin{figure}
	\centering
	\includegraphics[width=6cm,height=6cm]{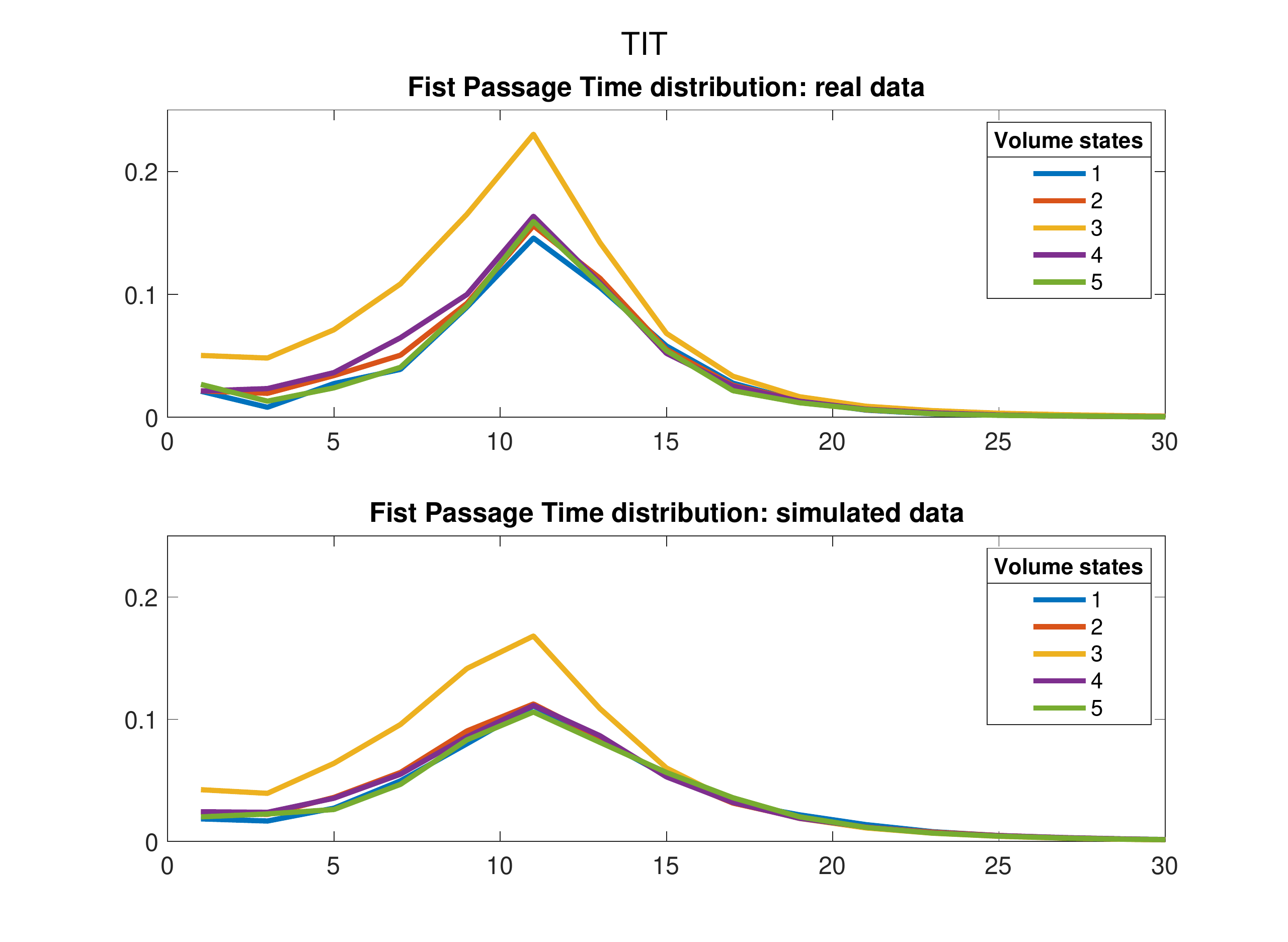}
	\includegraphics[width=6cm,height=6cm]{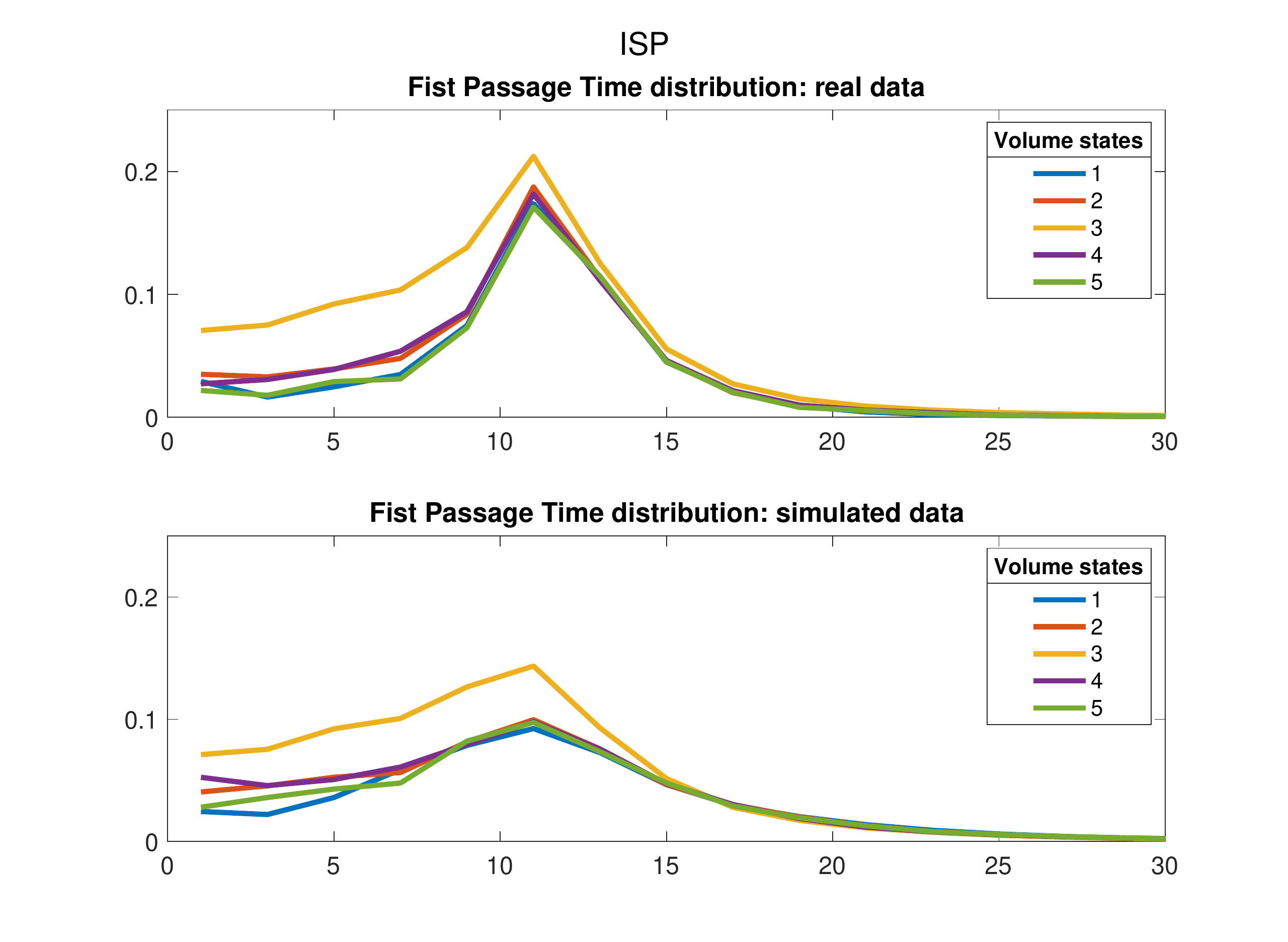}
	\caption{First passage time distribution of real data compared with synthetic data.} \label{fpt_vol_dip}
\end{figure}

To verify if the price fptd depends on volume values we estimated the fptd as a function of the value of initial condition on the discretized volume returns. In this way, for each initial $v(t)$ value, we obtain a different fptd. At the same time, we verified if the proposed model also keeps this dependence structure.
In Figure \ref{fpt_vol_dip} we show the results and comparison with real data (for two of the given stocks). We fixed a price increment threshold at $0.5\%$. It is worth noting that data are sampled at 1 minute, then it takes 10 to 15 minutes to reach this price change.

\begin{figure}
	\centering
	\includegraphics[width=8cm]{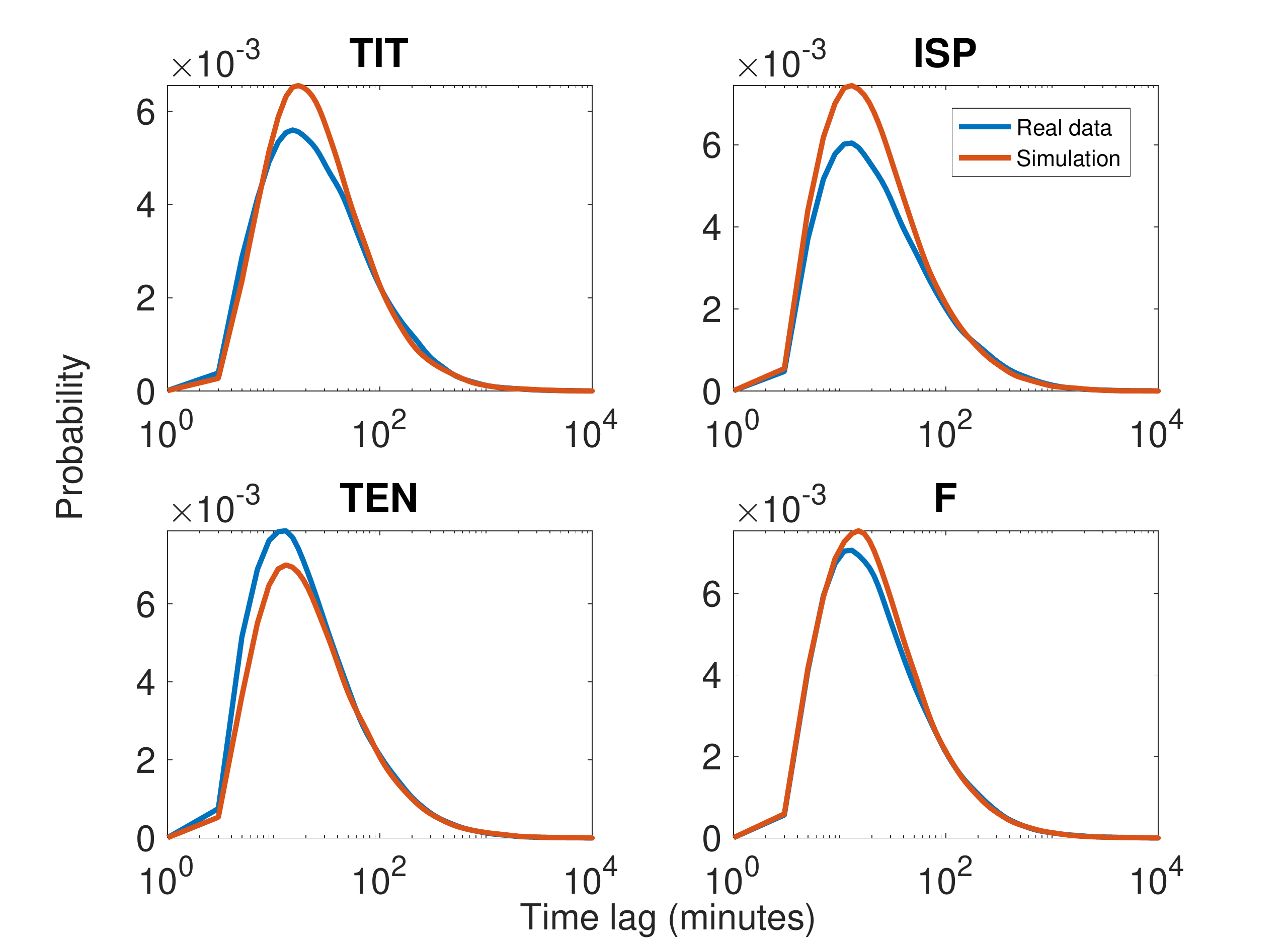}
	\caption{First passage time distribution of real data compared with synthetic data.} \label{fpt_cong}
\end{figure}

From the figure we can see that there is, indeed, a dependence on the initial conditions on volume. Furthermore, the dependence lasts up to 15 minutes and after this time all distributions converge to similar values. The behaviour depends on the different stocks but almost all of them show faster achievements of threshold when the initial volume return is in the third state.   
Figure \ref{fpt_vol_dip} also shows that, although with some differences, the model has the same behaviour of real data, there is a dependence on the initial conditions of volume return and it is very similar to real data.
Finally, in Figure \ref{fpt_cong} we show the joint first passage time distribution that represent the first time that both $S(t)$ and $V(t)$ cross a given threshold. Again, the price increment threshold is set at $0.5\%$ while the volume increment threshold is $100$. 

Overall we can say that the model is able to capture all statistical features of real data keeping all the dependencies between price, volumes and waiting times. Furthermore, we found very good agreement between real data and model also for the first passage time distribution.

\section{Conclusions}
In this work we have advanced a new stochastic model, based on Weigthed-Indexed Semi-Markov Chain, for modelling price, volumes and waiting times in high frequency finance. After showing, by analyzing real data, all the empirical evidences that support the use of a multivariate model, we defined the probabilistic structure of the model and give a detailed mathematical implementation.
Furthermore, mathematical expressions for covariance and first passage time distributions are given. In the last part we show, by using Monte Carlo simulations, that the model has the same statistical features of real data. In fact, the proposed model is able to reproduce the autocorrelation functions, the dependence between price and volume and the first passage time distributions. Further development can be the use of the model in portfolio optimization, development of risk measure and volatility forecasting.

\section{Appendix}
{\em Proof (of Lemma (\ref{homo}))}\\
Given the information set $(J,T)_{-m}^{n+1}=(i,t)_{-m}^{n+1}$ we can proceed to compute the value of the index process at the $(n+1)-{th}$ transition through formula $(\ref{funcrela})$ and assumption {\bf{A1}}:
\begin{equation}
\begin{aligned}
I_{n+1}^{J}(\lambda)&=\sum_{r=0}^{m+n+1-1}\sum_{a=T_{n+1-1-r}}^{T_{n+1-r}-1}f^{\lambda}(J_{n+1-1-r},T_{n+1}-a)+f^{\lambda}(J_{n+1},T_{n+1}-T_{n+1})
\\
& =\sum_{r=0}^{m+n}\sum_{a=t_{n-r}}^{t_{n+1-r}-1}f^{\lambda}(i_{n-r},t_{n+1}-a)+f^{\lambda}(i_{n+1},0).
\end{aligned}
\end{equation}
For simplicity of notation, denote by $x$ this value, i.e. $I_{n+1}^{J}(\lambda)=x$. Thus,
\begin{equation}
\begin{aligned}
& \mathbb{P}[J_{n+2}\leq j,\: T_{n+2}-T_{n+1}= t |(J,T)_{-m}^{n+1}=(i,t)_{-m}^{n+1}]\\
& =\mathbb{P}[J_{n+2}\leq j,\: T_{n+2}-T_{n+1}= t |J_{n+1}=i_{n+1},I_{n+1}^{J}(\lambda)=x]=q(i_{n+1},x;j,t).
\end{aligned}
\end{equation}
\indent Let us consider now the probability $\mathbb{P}[J_{n+1}\leq j,\: T_{n+1}-T_{n}= t |(J,T)_{-m-1}^{n}=\circ((i,t)_{-m}^{n+1})]$ and apply the definition of the shift operator to have:
\begin{equation}
\circ((i,t)_{-m}^{n+1})=(s,k)_{-m-1}^{n},
\end{equation}
and in turn
\begin{equation}
\begin{aligned}
& \mathbb{P}[J_{n+1}\leq j,\: T_{n+1}-T_{n}= t |(J,T)_{-m-1}^{n}=\circ((i,t)_{-m}^{n+1})]\\
& =\mathbb{P}[J_{n+1}\leq j,\: T_{n+1}-T_{n}= t |(J,T)_{-m-1}^{n}=(s,k)_{-m-1}^{n})]\\
& =\mathbb{P}[J_{n+1}\leq j,\: T_{n+1}-T_{n}= t |J_{n}=s_{n},I_{n}^{J}(\lambda)=b],
\end{aligned}
\end{equation}
where
\begin{equation}
b=\sum_{r=0}^{m+n}\sum_{a=k_{n-1-r}}^{k_{n-r}-1}f^{\lambda}(s_{n-1-r},k_{n}-a)+f^{\lambda}(s_{n},0).
\end{equation}
Since $s_{n-1-r}=i_{n-r}$ and $k_{n-1-r}=t_{n-r}-t_{n+1}$ it follows that 
$$
b=\sum_{r=0}^{m+n}\sum_{a=t_{n-r}-t_{n+1}}^{t_{n-r+1}-t_{n+1}-1}f^{\lambda}(i_{n-r},-a)+f^{\lambda}(i_{n+1},0).
$$
A change of variable $y=a+t_{n+1}$ gives
$$
b=\sum_{r=0}^{m+n}\sum_{y=t_{n-r}}^{t_{n-r+1}-1}f^{\lambda}(i_{n-r},t_{n+1}-y)+f^{\lambda}(i_{n+1},0)=x.
$$
Accordingly we get
\begin{equation*}
\begin{aligned}
& \mathbb{P}[J_{n+1}\leq j,\: T_{n+1}-T_{n}= t |(J,T)_{-m-1}^{n}=\circ((i,t)_{-m}^{n+1})]\\
& =\mathbb{P}[J_{n+1}\leq j,\: T_{n+1}-T_{n}= t |J_{n}=i_{n+1},I_{n}^{J}(\lambda)=x]=q(i_{n+1},x;j,t),
\end{aligned}
\end{equation*}
which completes the proof.\\

{\em Proof (of Theorem (\ref{kernelproof}))}\\
The kernel of the triplet process has been represented in formula $(\ref{otto})$ as follows:
$$
q^{JV}(\mathcal{A}_{n,s}^{JV};j,a,t)=\mathbb{P}[\tilde{J}_{n+1}\leq j,\: \tilde{V}_{n+1}\leq a | \mathcal{A}_{n,s}^{JVT}]\cdot {\mathbb{P}[\tilde{X}_{n}= t |\mathcal{A}_{n,s}^{JV}]},
$$
and from assumption {\bf{A3}} we get ${\tilde{h}_{i,v}(x,w;t)}={\mathbb{P}[\tilde{X}_{n}= t |\mathcal{A}_{n,s}^{JV}]}$.\\
\indent Thus, it remains to evaluate the conditional probability of the joint distribution of log-return and log-volume. Let consider the case when $j\geq 0$ and $a\geq 0$ and introduce the notation $F^{|J|}(j)$ and $F^{|V|}(a)$ to denote in a compact form the marginal distributions of the copula.\\
\indent Let us consider the following representation:
\begin{equation}
\label{4add}
\begin{aligned}
& \mathbb{P}[\tilde{J}_{n+1}\leq j,\: \tilde{V}_{n+1}\leq a | \mathcal{A}_{n,s}^{JVT}]=\mathbb{P}[|\tilde{J}_{n+1}|\leq j,\: |\tilde{V}_{n+1}|\leq a | \mathcal{A}_{n,s}^{JVT}]\\
& + \mathbb{P}[|\tilde{J}_{n+1}|> j,\: \eta_{n+1}^{J}=-1,\: |\tilde{V}_{n+1}|\leq a | \mathcal{A}_{n,s}^{JVT}]\\
& +\mathbb{P}[|\tilde{J}_{n+1}|> j,\: \eta_{n+1}^{J}=-1,\: |\tilde{V}_{n+1}|> a, \eta_{n+1}^{V}=-1 | \mathcal{A}_{n,s}^{JVT}]\\
& + \mathbb{P}[|\tilde{J}_{n+1}|\leq j,\: |\tilde{V}_{n+1}|> a, \eta_{n+1}^{V}=-1 | \mathcal{A}_{n,s}^{JVT}].
\end{aligned}
\end{equation}
\indent Let us proceed to the computation of each one of the four addenda in $(\ref{4add})$. From assumption {\bf{A4}} we know that
\begin{equation}
\label{1}
\mathbb{P}[|\tilde{J}_{n+1}|\leq j,\: |\tilde{V}_{n+1}|\leq a | \mathcal{A}_{n,s}^{JVT}]=\mathcal{C}\Big(F^{|J|}(j), F^{|V|}(a)\Big).
\end{equation}
\indent Next consider $\mathbb{P}[|\tilde{J}_{n+1}|> j,\: \eta_{n+1}^{J}=-1,\: |\tilde{V}_{n+1}|> a, \eta_{n+1}^{V}=-1 | \mathcal{A}_{n,s}^{JVT}]$. From assumption {\bf{A5}} this probability is equal to 
\begin{equation}
\label{2}
\begin{aligned}
& \mathbb{P}[\eta_{n+1}^{J}=-1]\cdot \mathbb{P}[\eta_{n+1}^{V}=-1]\cdot \mathbb{P}[|\tilde{J}_{n+1}|> j,\: |\tilde{V}_{n+1}|> a | \mathcal{A}_{n,s}^{JVT}]\\
& =(1-p^{J})(1-p^{V})\{1- \mathbb{P}[|\tilde{J}_{n+1}|\leq j,\: |\tilde{V}_{n+1}|\leq a | \mathcal{A}_{n,s}^{JVT}]\\
& -\mathbb{P}[|\tilde{J}_{n+1}|> j,\: |\tilde{V}_{n+1}|\leq a | \mathcal{A}_{n,s}^{JVT}] -\mathbb{P}[|\tilde{J}_{n+1}|\leq j,\: |\tilde{V}_{n+1}|> a | \mathcal{A}_{n,s}^{JVT}]\}\\
& =(1-p^{J})(1-p^{V})\{1-\mathcal{C}\Big(F^{|J|}(j), F^{|V|}(a)\Big)-(\mathbb{P}[|\tilde{J}_{n+1}|\leq j | \mathcal{A}_{n,s}^{JVT}]\\
& -\mathbb{P}[|\tilde{J}_{n+1}|\leq j,\: |\tilde{V}_{n+1}|\leq a | \mathcal{A}_{n,s}^{JVT}]) -(\mathbb{P}[|\tilde{V}_{n+1}|\leq a | \mathcal{A}_{n,s}^{JVT}]\\
& -\mathbb{P}[|\tilde{J}_{n+1}|\leq j, |\tilde{V}_{n+1}|\leq a | \mathcal{A}_{n,s}^{JVT}])\}\\
& =(1-p^{J})(1-p^{V})\Big\{1-\mathcal{C}\Big(F^{|J|}(j), F^{|V|}(a)\Big)-F^{|J|}(j) +  \mathcal{C}\Big(F^{|J|}(j), F^{|V|}(a)\Big)\\
& - F^{|V|}(a) +  \mathcal{C}\Big(F^{|J|}(j), F^{|V|}(a)\Big)\Big\}\\
& =(1-p^{J})(1-p^{V})[1-F^{|J|}(j) - F^{|V|}(j) +  \mathcal{C}\Big(F^{|J|}(j), F^{|V|}(a)\Big)].
\end{aligned}
\end{equation}
\indent Then, proceed to compute $\mathbb{P}[|\tilde{J}_{n+1}|> j,\: \eta_{n+1}^{J}=-1,\: |\tilde{V}_{n+1}|\leq a | \mathcal{A}_{n,s}^{JVT}]$. Apply again assumptions {\bf{A4}} and {\bf{A5}} to get 
\begin{equation}
\label{3}
\begin{aligned}
& \mathbb{P}[\eta_{n+1}^{J}=-1]\cdot \mathbb{P}[|\tilde{J}_{n+1}|> j,\: |\tilde{V}_{n+1}|\leq a | \mathcal{A}_{n,s}^{JVT}]\\
& =(1-p^{J})[\mathbb{P}[|\tilde{V}_{n+1}|\leq a | \mathcal{A}_{n,s}^{JVT}]-\mathbb{P}[|\tilde{J}_{n+1}|\leq j, |\tilde{V}_{n+1}|\leq a | \mathcal{A}_{n,s}^{JVT}]]\\
& =(1-p^{J})\Big[F^{|V|}(a) -  \mathcal{C}\Big(F^{|J|}(j), F^{|V|}(a)\Big)\Big].
\end{aligned}
\end{equation}
Analogous computations allow to get
\begin{equation}
\label{4}
\begin{aligned}
& \mathbb{P}[|\tilde{J}_{n+1}|\leq j,\: \eta_{n+1}^{V}=-1,\: |\tilde{V}_{n+1}|> a | \mathcal{A}_{n,s}^{JVT}]\\
& =(1-p^{V})\Big[F^{|J|}(j) -  \mathcal{C}\Big(F^{|J|}(j), F^{|V|}(a)\Big)\Big].
\end{aligned}
\end{equation}
\indent A substitution of $(\ref{1})$, $(\ref{2})$, $(\ref{3})$ and $(\ref{4})$ into $(\ref{4add})$ and some algebraic manipulations produces
\begin{equation}
\label{quasi}
\begin{aligned}
&\mathbb{P}[\tilde{J}_{n+1}\leq j,\: \tilde{V}_{n+1}\leq a | \mathcal{A}_{n,s}^{JVT}]=\\
& 1-p^{J}p^{V}\cdot \Big(1-F^{|J|}(j)-F^{|V|}(a)+\mathcal{C}\big(F^{|J|}(j), F^{|V|}(a)\big)\Big)\\
& -p^{V}(1-F^{|V|}(a))-p^{J}(1-F^{|J|}(j)).
\end{aligned}
\end{equation} 
A multiplication of $(\ref{quasi})$ by ${\tilde{h}_{i,v}(x,w;t)}$ concludes the proof for the case (i). The remaining cases (ii) - (iv) can be accomplished by similar arguments.\\

{\em Proof (of Theorem (\ref{prop}))}\\
\begin{equation*}
R_{(\rho ; \psi)}((i,v,t)_{-m}^{0},u;t)=\mathbb{P}[\Gamma_{(\rho ; \psi)} >t|(\tilde{J},\tilde{V},\tilde{T})_{-m}^{0}=(i,v,t)_{-m}^{0},\tilde{B}(u)=u]
\end{equation*}
\begin{equation}
\label{a1}
= \mathbb{P}[\Gamma_{(\rho ; \psi)} >t, \tilde{T}_{1}>t|(\tilde{J},\tilde{V},\tilde{T})_{-m}^{0}=(i,v,t)_{-m}^{0},\tilde{B}(u)=u]
\end{equation}
\begin{equation}
\label{a2}
+\mathbb{P}[\Gamma_{(\rho ; \psi)} >t, \tilde{T}_{1}\leq t|(\tilde{J},\tilde{V},\tilde{T})_{-m}^{0}=(i,v,t)_{-m}^{0},\tilde{B}(u)=u].
\end{equation}
\indent By the definition of conditional probability $(\ref{a1})$ can be written as
\begin{equation}
\label{a3}
\begin{aligned}
& \mathbb{P}[\Gamma_{(\rho ; \psi)} >t |\tilde{T}_{1}>t, (\tilde{J},\tilde{V},\tilde{T})_{-m}^{0}=(i,v,t)_{-m}^{0},\tilde{B}(u)=u]\\
& \times \mathbb{P}[\tilde{T}_{1}>t|(\tilde{J},\tilde{V},\tilde{T})_{-m}^{0}=(i,v,t)_{-m}^{0},\tilde{B}(u)=u].
\end{aligned}
\end{equation}
\indent Note that by definition $\tilde{X}_{0}:=\tilde{T}_{1}-\tilde{T}_{0}$ but since $\tilde{T}_{0}=t_{0}=0$, we can replace $\tilde{T}_{1}$ with the corresponding sojourn time $\tilde{X}_{0}$. Also note that the event $\{\tilde{B}(u)=u\}$ is equivalent to the event $\{\tilde{T}_{N(u)}=0, \tilde{T}_{N(u)+1}>u\}$. The latter equality between events means that at least one between returns and volumes did last transition at time $t_{0}=0$ and the other process made its last transition at some time before. Let $b^{J}$ and $b^{V}$ generically denote the times since last transition of the backward recurrence time processes, i.e.
\begin{equation*}
\tilde{T}_{0}-{T}_{0}^{J}=b^{J},\,\,\,\tilde{T}_{0}-{T}_{0}^{V}=b^{V}.
\end{equation*}
\indent Besides, note that the information set $(\tilde{J},\tilde{V},\tilde{T})_{-m}^{0}=(i,v,t)_{-m}^{0}$ generates a value of the index process of returns equal to
\begin{equation}
\tilde{I}_{0}^{V}=\sum_{r=0}^{m-1}\sum_{a=t_{-r-1}}^{t_{-r}-1}f^{\lambda_{J}}(i_{-r-1},-a)+f^{\lambda_{J}}(i_{0},0)=:\alpha_{0},
\end{equation}
and of the index process of volumes equal to
\begin{equation}
\tilde{I}_{0}^{V}=\sum_{r=0}^{m-1}\sum_{a=t_{-r-1}}^{t_{-r}-1}g^{\lambda_{V}}(v_{-r-1},-a)+g^{\lambda_{V}}(v_{0},0)=:\beta_{0}.
\end{equation}
\indent Thus, in virtue of assumption {\bf{A2}}, the probability $(\ref{a3})$ becomes equal to
\begin{equation}
\label{a4}
\begin{aligned}
& \mathbb{P}[\Gamma_{(\rho ; \psi)} >t |\tilde{X}_{0}>t, \tilde{J}_{0}=i_{0}, \tilde{V}_{0}=v_{0}, \tilde{I}_{0}^{J}=\alpha_{0}, \tilde{I}_{0}^{V}=\beta_{0}, \tilde{T}_{1}>u, \tilde{T}_{0}-{T}_{0}^{J}=b^{J},\tilde{T}_{0}-{T}_{0}^{V}=b^{V}]\\
& \times \mathbb{P}[\tilde{X}_{0}>t|(\tilde{J},\tilde{V},\tilde{T})_{-m}^{0}=(i,v,t)_{-m}^{0},\tilde{B}(u)=u].
\end{aligned}
\end{equation}
\indent Nevertheless, according to assumption {\bf{A3}}, we have
\begin{equation*}
\mathbb{P}[\tilde{X}_{0}>t|(\tilde{J},\tilde{V},\tilde{T})_{-m}^{0}=(i,v,t)_{-m}^{0},\tilde{B}(u)=u]
\end{equation*}
\begin{equation*}
=\mathbb{P}[\tilde{X}_{0}>t|\tilde{X}_{0}>u, \tilde{J}_{0}=i_{0}, \tilde{V}_{0}=v_{0}, \tilde{I}_{0}^{J}=\alpha_{0}, \tilde{I}_{0}^{V}=\beta_{0}]
\end{equation*}
\begin{equation}
\label{a5}
=\frac{\mathbb{P}[\tilde{X}_{0}>t|\tilde{J}_{0}=i_{0}, \tilde{V}_{0}=v_{0}, \tilde{I}_{0}^{J}=\alpha_{0}, \tilde{I}_{0}^{V}=\beta_{0}]}{\mathbb{P}[\tilde{X}_{0}>u| \tilde{J}_{0}=i_{0}, \tilde{V}_{0}=v_{0}, \tilde{I}_{0}^{J}=\alpha_{0}, \tilde{I}_{0}^{V}=\beta_{0}]}=\frac{1-\tilde{H}_{i_{0},v_{0}}(\alpha_{0},\beta_{0};t)}{1-\tilde{H}_{i_{0},v_{0}}(\alpha_{0},\beta_{0};u)}.
\end{equation}
\indent By the definition of joint first passage time we have that
\begin{equation*}
\mathbb{P}[\Gamma_{(\rho ; \psi)} >t |\tilde{X}_{0}>t, \tilde{J}_{0}=i_{0}, \tilde{V}_{0}=v_{0}, \tilde{I}_{0}^{J}=\alpha_{0}, \tilde{I}_{0}^{V}=\beta_{0}, \tilde{T}_{1}>u, \tilde{T}_{0}-{T}_{0}^{J}=b^{J},\tilde{T}_{0}-{T}_{0}^{V}=b^{V}]
\end{equation*} 
\begin{equation}
\label{a6}
=\mathbb{P}[\min\{\tau \geq 0: \{\tilde{M}_{0}^{J}(\tau)\geq \rho\} \cup \{\tilde{M}_{0}^{V}(\tau)\geq \psi\}\} | \mathcal{A}_{0,0}^{JV}, \tilde{T}_{1}>u],
\end{equation} 
\noindent where 
$$\mathcal{A}_{0,0}^{JV}=\{\tilde{J}_{0}=i_{0}, \tilde{V}_{0}=v_{0}, \tilde{I}_{0}^{J}=\alpha_{0}, \tilde{I}_{0}^{V}=\beta_{0}, \tilde{T}_{0}=0, \tilde{T}_{0}-{T}_{0}^{J}=b^{J},\tilde{T}_{0}-{T}_{0}^{V}=b^{V}\}.$$
\indent It is clear that since $i_{0}\geq 0$, $v_{0}\geq 0$ and $\tilde{T}_{1}>t$, the processes  $\tilde{M}_{0}^{J}(\tau)$ and $\tilde{M}_{0}^{V}(\tau)$ are both increasing with respect to the variable $\tau$. Accordingly, 
\[
\max_{\tau\in \{0,1,\ldots,t\}}\{\tilde{M}_{0}^{J}(\tau)\}=\tilde{M}_{0}^{J}(t)=e^{\sum_{r=0}^{t-1}\tilde{Z}^{J}(r)}=e^{i_{0}t},
\] 
and analogously $\max_{\tau\in \{0,1,\ldots,t\}}\{\tilde{M}_{0}^{V}(\tau)\}=e^{v_{0}t}$. Thus, formula $(\ref{a6})$ becomes 
\begin{equation}
\label{a7}
\mathbb{P}[e^{i_{0}t}< \rho, e^{v_{0}t}< \psi | \mathcal{A}_{0,0}^{JV}, \tilde{T}_{1}>u]=1_{\{e^{i_{0}t}< \rho\}}1_{\{e^{v_{0}t}< \psi\}}.
\end{equation} 
\indent A substitution of $(\ref{a7})$ and $(\ref{a5})$ in $(\ref{a4})$ gives:
\begin{equation}
\label{pezzo1}
\begin{aligned}
&\mathbb{P}[\Gamma_{(\rho ; \psi)} >t|(\tilde{J},\tilde{V},\tilde{T})_{-m}^{0}=(i,v,t)_{-m}^{0},\tilde{B}(u)=u]\\
& =1_{\{e^{i_{0}t}< \rho\}}1_{\{e^{v_{0}t}< \psi\}}\frac{1-\tilde{H}_{i_{0},v_{0}}(\alpha_{0},\beta_{0};t)}{1-\tilde{H}_{i_{0},v_{0}}(\alpha_{0},\beta_{0};u)}.
\end{aligned}
\end{equation}
\indent It remains to compute probability $(\ref{a2})$. By the law of total probability and by the definition of conditional probability we have the following chain of equality:
\begin{equation*}
\mathbb{P}[\Gamma_{(\rho ; \psi)} >t, \tilde{T}_{1}\leq t|(\tilde{J},\tilde{V},\tilde{T})_{-m}^{0}=(i,v,t)_{-m}^{0},\tilde{B}(u)=u]= \sum_{t_{1}=1}^{t}\! \int_{-\infty}^{+\infty}\!\!\! \int_{-\infty}^{+\infty}\!\!\!\!\!\mathbb{P}[\Gamma_{(\rho ; \psi)} >t,
\end{equation*}
\begin{equation*}
, \tilde{T}_{1}=t_{1}, \tilde{J}_{1}\in (i_{1},i_{1}+di_{1}),\tilde{V}_{1}\in (v_{1},v_{1}+dv_{1})|(\tilde{J},\tilde{V},\tilde{T})_{-m}^{0}=(i,v,t)_{-m}^{0},\tilde{B}(u)=u]
\end{equation*}
\begin{equation*}
\begin{aligned}
&= \sum_{t_{1}=1}^{t}\int_{-\infty}^{+\infty}\int_{-\infty}^{+\infty}\mathbb{P}[\Gamma_{(\rho ; \psi)} >t | \tilde{T}_{1}=t_{1}, \tilde{J}_{1}=i_{1}, \tilde{V}_{1}=v_{1}, (\tilde{J},\tilde{V},\tilde{T})_{-m}^{0}=(i,v,t)_{-m}^{0},\tilde{B}(u)=u]\\
& \times  \mathbb{P}[\tilde{T}_{1}=t_{1}, \tilde{J}_{1}\in (i_{1},i_{1}+di_{1}), \tilde{V}_{1}\in (v_{1},v_{1}+dv_{1}) | (\tilde{J},\tilde{V},\tilde{T})_{-m}^{0}=(i,v,t)_{-m}^{0},\tilde{B}(u)=u].
\end{aligned}
\end{equation*}
\indent Let start to compute the following probability:
\begin{equation*}
\mathbb{P}[\tilde{T}_{1}=t_{1}, \tilde{J}_{1}\in (i_{1},i_{1}+di_{1}), \tilde{V}_{1}\in (v_{1},v_{1}+dv_{1}) | (\tilde{J},\tilde{V},\tilde{T})_{-m}^{0}=(i,v,t)_{-m}^{0},\tilde{B}(u)=u]
\end{equation*}
\begin{equation*}
=\mathbb{P}[\tilde{T}_{1}=t_{1}, \tilde{J}_{1}\in (i_{1},i_{1}+di_{1}), \tilde{V}_{1}\in (v_{1},v_{1}+dv_{1}) | \mathcal{A}_{0,0}^{JV},\tilde{T}_{1}>u]
\end{equation*}
\begin{equation*}
=\frac{\mathbb{P}[u<\tilde{T}_{1}=t_{1}, \tilde{J}_{1}\in (i_{1},i_{1}+di_{1}), \tilde{V}_{1}\in (v_{1},v_{1}+dv_{1}) | \mathcal{A}_{0,0}^{JV}]}{\mathbb{P}[\tilde{T}_{1}>u  | \mathcal{A}_{0,0}^{JV}]}
\end{equation*}
\begin{equation}
\label{b1}
=\frac{1_{\{t_{1}>u\}}\frac{\partial^{2}q^{JV}(i_{0},v_{0},\alpha_{0},\beta_{0};i_{1},v_{1},t_{1})}{\partial i_{1} \partial v_{1}}di_{1}dv_{1}}{1-\tilde{H}_{i_{0},v_{0}}(\alpha_{0},\beta_{0};u)}.
\end{equation}
\indent It remains to compute
\begin{equation*}
\mathbb{P}[\Gamma_{(\rho ; \psi)} >t | \tilde{T}_{1}=t_{1}, \tilde{J}_{1}=i_{1}, \tilde{V}_{1}=v_{1}, (\tilde{J},\tilde{V},\tilde{T})_{-m}^{0}=(i,v,t)_{-m}^{0},\tilde{B}(u)=u]
\end{equation*}
\begin{equation}
\label{b2}
=\mathbb{P}[\max_{s\in \{0,1,\ldots,t\}}\{\tilde{M}_{0}^{J}(s)\}<\rho, \max_{s\in \{0,1,\ldots,t\}}\{\tilde{M}_{0}^{V}(s)\}<\psi | (\tilde{J},\tilde{V},\tilde{T})_{-m}^{1}=(i,v,t)_{-m}^{1},\tilde{B}(u)=u].
\end{equation}
\indent Now observe that since $\tilde{T}_{1}=t_{1}$ we have that
\begin{equation*}
\max_{s\in \{0,1,\ldots,t\}}\{\tilde{M}_{0}^{J}(s)\}=\max \{\max_{s\in \{0,1,\ldots,t_{1}\}}\{\tilde{M}_{0}^{J}(s)\}, \max_{s\in \{1,\ldots,t-t_{1}\}}\{\tilde{M}_{0}^{J}(t_{1}+s)\}\},
\end{equation*}
 and due to the fact that $i_{0}\geq 0$ it results that $\max_{s\in \{0,1,\ldots,t_{1}\}}\{\tilde{M}_{0}^{J}(s)\}=\tilde{M}_{0}^{J}(t_{1})=e^{t_{1}i_{0}}$. Accordingly we can deduce that
\begin{equation*}
\max_{s\in \{0,1,\ldots,t\}}\{\tilde{M}_{0}^{J}(s)\}=\max \{e^{t_{1}i_{0}}, \max_{s\in \{1,\ldots,t-t_{1}\}} \{e^{t_{1}i_{0}}e^{\sum_{r=0}^{s-1}\tilde{Z}^{J}(t_{1}+r)}\}\}.
\end{equation*}
\indent Similarly we have 
\begin{equation*}
\max_{s\in \{0,1,\ldots,t\}}\{\tilde{M}_{0}^{V}(s)\}=\max \{e^{t_{1}v_{0}}, \max_{s\in \{1,\ldots,t-t_{1}\}} \{e^{t_{1}v_{0}}e^{\sum_{r=0}^{s-1}\tilde{Z}^{V}(t_{1}+r)}\}\}.
\end{equation*}
\indent Thus by substitution, the probability $(\ref{b2})$ becomes
\begin{equation*}
\begin{aligned}
& =\mathbb{P}\Big[\max \Big\{e^{t_{1}i_{0}}, \max_{s\in \{1,\ldots,t-t_{1}\}} \{e^{t_{1}i_{0}}e^{\sum_{r=0}^{s-1}\tilde{Z}^{J}(t_{1}+r)}\}\Big\}<\rho, \\
& ,\max \Big\{e^{t_{1}v_{0}}, \max_{s\in \{1,\ldots,t-t_{1}\}} \{e^{t_{1}v_{0}}e^{\sum_{r=0}^{s-1}\tilde{Z}^{V}(t_{1}+r)}\}\Big\}<\psi | (\tilde{J},\tilde{V},\tilde{T})_{-m}^{1}=(i,v,t)_{-m}^{1},\tilde{B}(u)=u\Big].
\end{aligned}
\end{equation*}
\begin{equation*}
\begin{aligned}
&=1_{\{e^{t_{1}i_{0}}<\rho\}}1_{\{e^{t_{1}v_{0}}<\psi\}}\\
&\cdot \mathbb{P}[\max_{s\in \{1,\ldots,t-t_{1}\}} \{e^{t_{1}i_{0}}e^{\sum_{r=0}^{s-1}\tilde{Z}^{J}(t_{1}+r)}\}<\rho,  \max_{s\in \{1,\ldots,t-t_{1}\}} \{e^{t_{1}v_{0}}e^{\sum_{r=0}^{s-1}\tilde{Z}^{V}(t_{1}+r)}\}<\psi \\
& | (\tilde{J},\tilde{V},\tilde{T})_{-m}^{1}=(i,v,t)_{-m}^{1},\tilde{B}(u)=u].
\end{aligned}
\end{equation*}
\begin{equation*}
\begin{aligned}
& =1_{\{e^{t_{1}i_{0}}<\rho\}}1_{\{e^{t_{1}v_{0}}<\psi\}}\cdot \mathbb{P}[\max_{s\in \{1,\ldots,t-t_{1}\}} \{e^{\sum_{r=0}^{s-1}\tilde{Z}^{J}(t_{1}+r)}\}<\frac{\rho}{e^{t_{1}i_{0}}},\\
& ,\max_{s\in \{1,\ldots,t-t_{1}\}} \{e^{\sum_{r=0}^{s-1}\tilde{Z}^{V}(t_{1}+r)}\}<\frac{\psi}{e^{t_{1}v_{0}}} | (\tilde{J},\tilde{V},\tilde{T})_{-m}^{1}=(i,v,t)_{-m}^{1},\tilde{B}(u)=u].
\end{aligned}
\end{equation*}
\begin{equation*}
=1_{\{e^{t_{1}i_{0}}<\rho\}}1_{\{e^{t_{1}v_{0}}<\psi\}}\cdot \mathbb{P}[\Gamma_{(\frac{\rho}{e^{t_{1}i_{0}}} ; \frac{\psi}{e^{t_{1}v_{0}}})} >t-t_{1} | (\tilde{J},\tilde{V},\tilde{T})_{-m}^{1}=(i,v,t)_{-m}^{1},\tilde{B}(u)=u].
\end{equation*}
\indent The latter probability, making use of definition \ref{sequence3} can be expressed as
\begin{equation*}
1_{\{e^{t_{1}i_{0}}<\rho\}}1_{\{e^{t_{1}v_{0}}<\psi\}}\cdot \mathbb{P}[\Gamma_{(\frac{\rho}{e^{t_{1}i_{0}}} ; \frac{\psi}{e^{t_{1}v_{0}}})} >t-t_{1} | (\tilde{J},\tilde{V},\tilde{T})_{-m-1}^{0}=\circ (i,v,t)_{-m}^{1},\tilde{B}(u)=u]
\end{equation*}
\begin{equation}
\label{b3}
=1_{\{e^{t_{1}i_{0}}<\rho\}}1_{\{e^{t_{1}v_{0}}<\psi\}}\cdot R_{\Big(\frac{\rho}{e^{i_{0}t_{1}}};\frac{\psi}{e^{v_{0}t_{1}}}\Big)}(\circ((i,v,t)_{-m}^{1}),0;t-t_{1}).
\end{equation}
\indent A substitution of $(\ref{b3})$ in $(\ref{b2})$ and then of the obtained quantity in $(\ref{a2})$ togheter with $(\ref{b1})$ concludes the proof.

\bibliographystyle{spmpsci}
\bibliography{references_GF}

\end{document}